\documentclass[pra,aps,superscriptaddress,showkeys,twocolumn,floats,floatfix,nobibnotes,amsfonts,amsmath,amssymb,longbibliography]{revtex4-1}
\usepackage{natbib}

\usepackage[utf8]{inputenc}
\usepackage[english]{babel}
\usepackage{commath}
\usepackage{siunitx}
\usepackage{tabularx}
\usepackage{xfrac}
\usepackage{gensymb}
\usepackage[normalem]{ulem} 

\usepackage{graphicx}

\usepackage[colorlinks=true,allcolors=blue]{hyperref}
\usepackage{color}
\usepackage{newtxmath}

\newcommand{\vecc}[1]{\mathbf{#1}} 

\newcommand{\SC}{superconducting}
\newcommand{\DOFF}{degrees of freedom}
\newcommand{\AHC}{AHC}
\newcommand{\ahc}{anti-Helmholtz coil}
\newcommand{\dlp}{double-loop}
\newcommand{\Dlp}{Double-loop}
\newcommand{\DLP}{DL}

\newcommand{\eref}[1]{equation \eqref{#1}}
\newcommand{\Eref}[1]{Equation \eqref{#1}}
\newcommand{\fref}[1]{figure \ref{#1}}
\newcommand{\Fref}[1]{Figure \ref{#1}}
\newcommand{\sref}[1]{section \ref{#1}}

\newcommand{\tref}[1]{table \ref{#1}}
\newcommand{\Tref}[1]{Table \ref{#1}}
\newcommand{\aref}[1]{appendix \ref{#1}}

\newcommand{\xtxt}[1]{{\typeout{#1}}}
\newcommand{\ntxt}[1]{{#1}}

\begin{document}

\title{Chip-based superconducting traps for levitation of micrometer-sized particles in the Meissner state}

\author{Martí Gutierrez Latorre}
\affiliation{Department of Microtechnology and Nanoscience (MC2), Chalmers University of Technology, Kemiv\"agen 9, SE-412 96 Gothenburg, Sweden}

\author{Joachim Hofer}
\affiliation{Vienna Center for Quantum Science and Technology (VCQ), Faculty of Physics, University of Vienna, Boltzmanngasse 5, Vienna, A-1090, Austria}

\author{Matthias Rudolph}
\affiliation{Department of Microtechnology and Nanoscience (MC2), Chalmers University of Technology, Kemiv\"agen 9, SE-412 96 Gothenburg, Sweden}

\author{Witlef Wieczorek}
\email{witlef.wieczorek@chalmers.se}
\affiliation{Department of Microtechnology and Nanoscience (MC2), Chalmers University of Technology, Kemiv\"agen 9, SE-412 96 Gothenburg, Sweden}

\begin{abstract}
We present a detailed analysis of two chip-based superconducting trap architectures capable of levitating micrometer-sized superconducting particles in the Meissner state. These architectures are suitable for performing novel quantum experiments with more massive particles or for force and acceleration sensors of unprecedented sensitivity. We focus in our work on a chip-based \ahc{}-type trap (\AHC{}) and a planar \dlp{} (\DLP{}) trap. We demonstrate their fabrication from superconducting Nb films and the fabrication of superconducting particles from Nb or Pb. We apply finite element modeling (FEM) to analyze these two trap architectures in detail with respect to trap stability and frequency. Crucially, in FEM we account for the complete three-dimensional geometry of the traps, finite magnetic field penetration into the levitated superconducting particle, demagnetizing effects, and flux quantization. We can, thus, analyze trap properties beyond assumptions made in analytical models. We find that realistic \AHC{} traps yield trap frequencies well above 10\,kHz for levitation of micrometer-sized particles and can be fabricated with a three-layer process, while \DLP{} traps enable trap frequencies below 1\,kHz and are simpler to fabricate in a single-layer process. Our numerical results guide future experiments aiming at levitating micrometer-sized particles in the Meissner state with chip-based superconducting traps. The modeling we use is also applicable in other scenarios using superconductors in the Meissner state, such as for designing superconducting magnetic shields or for calculating filling factors in superconducting resonators.
\end{abstract}

\maketitle

\section{Introduction}

Superconducting magnetic levitation \cite{moon_superconducting_1994,Brandt} is a fascinating phenomenon. Its applications range from demonstration experiments \cite{arkadiev_floating_1947} to precise measurements of gravity using the superconducting gravimeter \cite{goodkind_superconducting_1999}. Recently, theoretical proposals suggest the use of superconducting magnetic levitation as a means to enable new experiments in the field of quantum optics  \cite{SClevitation,ringlevitation}. Specifically, micrometer-sized superconducting or magnetic particles levitated by magnetic fields are proposed to lead to a new generation of quantum experiments that enable spatial superposition states of levitated particles \cite{SClevitation,ringlevitation,johnsson_macroscopic_2016,pino_-chip_2018}, or ultra-high sensitivities for measurement of forces or accelerations \cite{johnsson_macroscopic_2016,SCmagnetlevitation,jackson_kimball_precessing_2016}, with recent experiments along these lines \cite{slezak_cooling_2018,wang_dynamics_2019,timberlake_acceleration_2019,vinante_ultralow_2020,gieseler_single-spin_2020,zheng_room_2020}.

We consider levitation of superconducting particles in the Meissner state, inspired by Refs.\cite{SClevitation,ringlevitation,pino_-chip_2018}. Their stable levitation requires traps that generate a local magnetic field minimum accompanied by a field gradient \cite{frog}. Superconducting chip-based trap structures have already been developed in the context of atom optics for trapping atomic clouds on top of superconducting chips \cite{nirrengarten_realization_2006,fortagh_magnetic_2007,dikovsky_superconducting_2009,bernon_manipulation_2013}. However, in contrast to trapped atomic clouds, a levitated particle has a finite extent and, thus, requires accounting for its volume and the finite magnetic field penetration in the levitated object such that trap properties can be accurately predicted. 
Analytical formulas exist for idealized geometries, such as for levitation of a perfect diamagnetic sphere in a quadrupole field \cite{SClevitation} or in a field of four parallel wires \cite{pino_-chip_2018}, for a superconducting sphere in a quadrupole field \cite{hofer_analytic_2019}, for a perfect diamagnetic ring in a quadrupole field \cite{carles_2020} or can be derived for symmetric geometries and perfect diamagnetic objects using the image method \cite{lin_theoretical_2006}. However, in the general case when considering realistic three-dimensional trap geometries with reduced symmetry, trap wires of finite extent or arbitrary shapes of the levitated particle, analytical formulas do not exist and one has to resort to modeling using finite-element methods (FEM).

In our work, we present the fabrication and modeling of two promising chip-based trap architectures suitable for levitation of micrometer-sized superconducting objects of spherical, cylindrical or ring shape. We focus on multi-layer \ahc{}-like traps (\AHC{}) and  single-layer \dlp{} traps (\DLP{}). We first demonstrate fabrication of the traps using thin films of Nb \cite{pekola} and of particles made from Nb or Pb of spherical, cylindrical or ring shape. We then use FEM-based simulations to numerically calculate crucial trap parameters, such as stability, frequency and levitation height, for realistic geometries incorporating the finite extent of the wires and the non-symmetry of the traps.

\begin{figure}[t!bhp]
  \centering
  \includegraphics[width=0.9\linewidth]{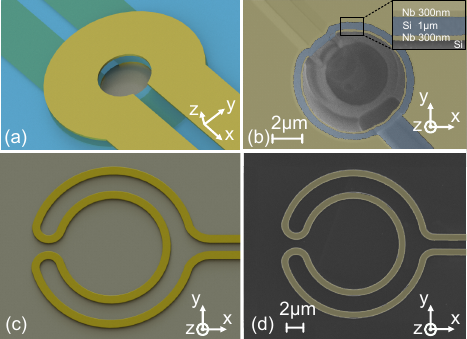}
\caption{Schematic drawings of (a) an \ahc{} (\AHC{}) trap and (c) a \dlp{} (\DLP{}) trap. Scanning electron microscope (SEM) images showing (b) the top view of the \AHC{} trap center, where the particle will levitate. The inset in (b) shows the cross section of the three-layer \AHC{} trap. (d) SEM image of a microfabricated \DLP{} trap.}
\label{fig:SEM}
\end{figure}

Our FEM simulations are based on implementing Maxwell-London equations in the static regime using the $\vecc{A}$-V formulation under the assumption that the levitated particles are in the Meissner state \cite{cordier_3-d_1999,cordier_finite-element_1999,grilli_finite-element_2005,campbell_introduction_2011}. We specifically assume levitation of a particle in the Meissner state, which has been proposed to minimize mechanical loss \cite{SClevitation,pino_-chip_2018}, a limiting factor for performing quantum experiments. We compare the numerical FEM results to idealized situations of increased symmetry, where analytical results can be obtained \cite{currentloop,lin_theoretical_2006,hofer_analytic_2019}. While the analytical results are indicative of the underlying physics, numerical modeling yields predictions independent of most idealizing assumptions. Finally, we apply FEM modeling to estimate the signal induced by the motion of a levitated particle in a nearby pick-up loop. This signal would be used to manipulate the center-of-mass motion of the particle in subsequent quantum experiments\cite{SClevitation}. 

\section{Microfabrication of traps and particles}\label{sec:fab}

In the following, we describe the microfabrication of chip-based traps from superconducting Nb films and of superconducting particles from Pb and Nb. Note that other superconducting materials, such as Al, can also be used. The choice of material determines the maximal allowed temperature of the cryogenic environment. While Pb and Nb, for example, allow levitation at liquid He temperatures, Al requires temperatures below 1.2\,K. Further, the particles need to be in the Meissner state to avoid mechanical loss \cite{SClevitation,pino_-chip_2018}. Hence, the magnetic field close to the particle surface must be smaller than the first critical field of the chosen material.

\subsection{Fabrication of traps}

\begin{figure}[t!bhp]
  \centering
  \includegraphics[width=\linewidth]{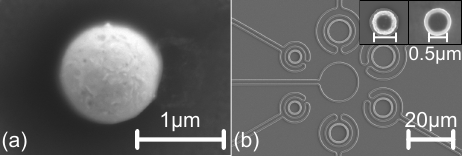}
   \caption{(a) SEM image of a Pb sphere on a Si substrate. (b) Several \DLP{} traps of different dimensions each with an in-situ fabricated Nb ring. The insets in (b) show a Nb cylinder and Nb ring fabricated on a Si substrate.}
    \label{fig:particles}
\end{figure}

The \AHC{}-type trap is formed by two coils arranged in an anti-Helmholtz-like configuration. This trap yields a large magnetic field gradient in the trap center, resulting in trap frequencies above 10\,kHz, see \sref{sec:ahc}. \Fref{fig:SEM}(a) shows a schematic and \fref{fig:SEM}(b) scanning electron microscope (SEM) images of a trap with 3$\,\muup$m inner coil radius and 1$\,\muup$m vertical coil separation fabricated in a three-layer process. The three layers used are Nb/Si/Nb, which are 300/1000/300\,nm thick, respectively. The lower Nb layer is sputtered first and subsequently patterned by optical lithography and etched using inductively coupled plasma-reactive ion etching (ICP-RIE) \cite{microfab}. Then, the Si layer is sputtered and subsequently etched via RIE to expose the contact pads of the lower Nb coil. The upper Nb layer is sputtered on top of this Si layer and structured. An electrical connection between the lower and upper Nb layer is facilitated by the Nb material sputtered on the sidewalls of the openings in the Si layer. Finally, a hole is etched through the three layers via ICP-RIE, which becomes the trapping region. 

An alternative trap arrangement consists of two concentric and co-planar coils that carry  counter-propagating currents.  A schematic of such a \DLP{} trap is shown in \fref{fig:SEM}(c), which can be regarded as a \AHC{}-type trap in the plane. \Fref{fig:SEM}(d) shows a microfabricated \DLP{} trap made from a 300\,nm thick Nb film and patterned via electron beam lithography (EBL). This trap generates a local energy minimum above the plane of the coils, where a particle will be stably levitated with trap frequencies below 1\,kHz, see \sref{sec:dltrap}. The \DLP{} trap has the advantages of a simple single-layer microfabrication process and that the trap region is not restricted by a vertical separation between coils like in the \AHC{}-trap.

We determined the properties of the 300\,nm thick Nb film from R-T, I-V and Hall effect measurements to have a $T_c\approx 9$\,K, a critical current density up to $j_c=5\cdot10^{11}$\,A/m$^2$ and a critical field $B_{c2}\approx 0.4$\,T, similar to previously reported values \cite{asada_superconductivity_1969,rusanov_depairing_2004,kim_critical_2009}. For the analysis of the traps, we will assume a current density in the coil wires of $1\cdot10^{11}$\,A/m$^2$ (unless otherwise stated), which is close to the measured critical current density.

\subsection{Fabrication of particles}

\begin{table*}[t!bhp]
    \centering
    \begin{tabular}{lllll}
    \hline\hline\\[-5pt]
     Model       &  Trap  & Particle  & Parameters & Comment \\[1.5pt]
     \hline\\[-5pt]
     
     Point particle \cite{currentloop} & 1D closed current loops & point particle & $I$, $r$ & point particle\\[1.5pt]
     
     Perfect diamagnet \cite{lin_theoretical_2006} & 1D closed current loops & superconducting sphere & $I$, $r$, $R$, $\lambda_L=0$ & image method \\[1.5pt]
     
     Superconducting sphere \cite{hofer_analytic_2019} & quadrupole field & superconducting sphere & $b$, $R$, $\lambda_L$ & sphere in Meissner state \\[1.5pt]

     FEM-2D-1D  & quasi-1D closed current loops  & rotationally symmetric & $I$, $r$, $[R_{2D}]$, $\lambda_L$ & 2D model with 1D wires\\
      &  (cross section $1\times 1$\,nm$^2$) &&&\\[1.5pt]

     FEM-2D & closed current loops & rotationally symmetric & $I$, $r$, $t$, $[R_{2D}]$, $\lambda_L$ & 2D model \\[1.5pt]
     
     FEM-3D & any shape & any shape & $I$, $[r_{3D}]$, $[R_{3D}]$, $\lambda_L$ & 3D model \\[1.5pt]
     \hline\hline
    \end{tabular}
    \caption{Different models we apply for calculating the trap architectures. The first three models are analytical models, while the other three are implemented in FEM. Parameters: current through wire $I$, magnetic field gradient at trap center  $b$, wire radius $r$, wire thickness $t$,  wire dimensions in 3D $[r_{3D}]$, sphere radius $R$, dimensions of rotational symmetric particle $[R_{2D}]$, particle dimensions in 3D $[R_{3D}]$, London penetration depth $\lambda_L$.}
    \label{tab:models}
\end{table*}

The particles can be obtained from particle powders or can be microfabricated directly in the trap. \Fref{fig:particles}(a) shows a spherical Pb particle individually selected from Pb powder. Note, however, that most particles in the powder are non-spherical and one has to pick-and-place the desired particles into the trap region. A systematic approach towards fabricating particles can rely on etching of thin superconducting layers. To this end, we fabricated cylinder- and ring-shaped particles directly on the trap chip by sputtering a 300\,nm thick Nb layer on top of a sacrificial layer of hard-baked resist, see  \fref{fig:particles}(b). The particle shape is patterned via EBL followed by ICP-RIE etching. The sacrificial resist layer is removed using oxygen plasma, releasing the particles onto the chip. 

\section{Numerical analysis of superconducting trap architectures}\label{sec:traps}

In the following, we systematically analyze the presented trap architectures with respect to the stability of the trap and achievable trap frequencies for different trap sizes and geometries of the levitated particle. Before we proceed with this analysis, we recall the conditions for achieving stable levitation and present the different models we are going to use.

\subsection{Models and assumptions}

Two requirements have to be met to achieve stable levitation \cite{Brandt,frog}, see the more detailed discussion in \aref{app:maglev}. First, the magnetic and gravitational force have to balance each other, such that the particle is levitated in free space above the chip surface. Second, the levitation position needs to be stable, i.e., the particle needs to experience a restoring force along each spatial direction. If these two conditions are met, we can calculate a trap frequency, $\omega_t$, from the gradient of the force, $F$, at the levitation position, $x_{\mathrm{lev}}$, via
\begin{equation}
    \left(\omega_t\right)^2=-\frac{1}{m}\frac{\partial F}{\partial x}\bigg\rvert_{x_{\mathrm{lev}}}\equiv-\frac{1}{m} k_t,
    \label{eq:trapfreq}
\end{equation}
where $m$ is the mass of the particle and $k_t$ is the spring constant of the trap. A non-spherical particle also requires rotational stability and, thus, we also analyze torques, $\vecc{\tau}_i$, rotating the particle around an axis $i$ by an angle $\theta_j$. If stable at $\theta_{\mathrm{lev}}$, we calculate a corresponding angular frequency, $\omega_i^{\tau}$, from
\begin{equation}
    \left(\omega_i^{\tau}\right)^2=-\frac{1}{I}\frac{\partial \tau_i}{\partial \theta_j}\bigg\rvert_{\theta_{\mathrm{lev}}}\equiv-\frac{1}{I} k_i^{\tau},
    \label{eq:restoringtorque}
\end{equation}
where $I$ is the moment of inertia of the particle and $k_i^{\tau}$ is the angular spring constant. \Eref{eq:trapfreq} and \eref{eq:restoringtorque} yield accurate trap frequencies as long as the force and torque depend linearly on displacement and angle, respectively, to which we restrict our analysis. Deviations can occur for larger particle amplitudes, see, e.g., Refs.~\cite{ricci_optically_2017,vinante_ultralow_2020}.

Knowing the magnetic field distribution of a particle in the trap allows calculating the necessary forces and torques, for details see \aref{app:maglev}. \Tref{tab:models} summarizes the analytical \cite{currentloop,lin_theoretical_2006,hofer_analytic_2019} and FEM models we use for calculating magnetic field distributions of the traps. We consider different levels of FEM modeling, which allow us comparing to the analytical models that necessarily make assumptions about the trap geometry or neglect the finite magnetic field penetration into the particle.

\begin{figure}[t!bhp]
  \centering
  \includegraphics[width=\linewidth]{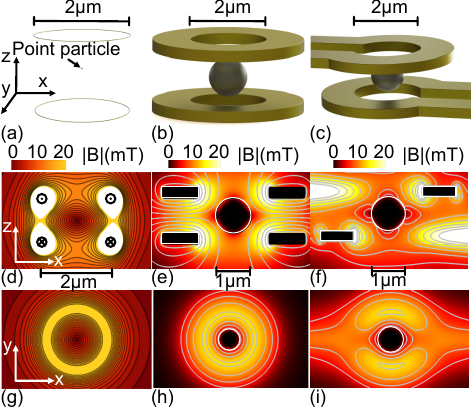}
   \caption{Geometry of an \AHC{} trap for calculations with the (a) point particle, (b) FEM-2D and (c) FEM-3D models. Magnetic field distributions for a cut in the XZ (d,e,f) and XY (g,h,i) planes that are tangent to the center of the trap. White areas show field strengths higher than the upper bound of the legend. Parameters used (cf.~Ref.~\cite{SClevitation}): inner coil diameter 2$\,\muup$m, outer coil diameter 4$\,\muup$m, coil separation 1$\,\muup$m, $I=30$\,mA in each coil, $R=0.5\,\muup$m superconducting sphere with $\lambda_{\textrm{L}}=50\,$nm, $\rho=8570\,$kg/m$^3$. From FEM-2D we obtain trap gradients of (9861, 9861, 19710)$\,$\si{T/m} along (x, y, z) in the center of an empty trap.}
    \label{fig:AHCtrap}
\end{figure}

The FEM modeling we use is based on the following assumptions. First, the particle is assumed to be in the Meissner state, which is motivated by the proposals of Refs.~\cite{SClevitation,pino_-chip_2018} and implemented in FEM via the $\vecc{A}$-$V$ formulation of the Maxwell-London equations \cite{cordier_3-d_1999,cordier_finite-element_1999,grilli_finite-element_2005,campbell_introduction_2011}, for details see \aref{app:FEMmodel}, for validation examples see \aref{app:validate} and for the FEM meshing see \aref{app:FEMmesh} (discretized using quadratic mesh discretization). We, thus, only consider trap fields that remain below the first critical field on the particle surface (we are restricting us to $B_c=0.08$\,T of Pb). Second, we account for flux quantization when considering levitation of a ring ad hoc by defining an area in the FEM model over which the flux should be constant. We neglect the flux in the interior of the material caused by the finite magnetic field penetration depth of the external field. This approximation is valid \cite{SCrings} for  $\Lambda/R \ll 1$ (we have $\Lambda/R<0.04$), where $\Lambda=\lambda_L^2/d$ is the two-dimensional effective penetration depth, $\lambda_L$ is the London penetration depth, $R$ is the lateral size of the superconducting object and $d$ its thickness. Third, for simplicity we model the wires as very low resistivity, diamagnetic normal conducting material carrying a uniform current across the wire geometry. The latter assumption is inspired by the situation of using a rectangular type-II superconducting film as wire material transporting a current under self-field that is close to its critical current density \cite{talantsev_current_2018}. Future extensions could model the wires using the critical state model \cite{bean_magnetization_1964,navau_macroscopic_2013,via_simultaneous_2014}, which would also allow analysis of various loss mechanisms \cite{grilli_computation_2014,mykola_v_2019}. Note that hysteresis or AC losses are negligible for the cases we are going to consider in \sref{sec:detect} \cite{SClevitation}. Finally, we need to consider that the magnetic field and the current density are gauge invariant. The gauge is fixed in the utilized FEM software COMSOL Multiphysics \cite{comsol} by implementing the Coulomb gauge at the cost of adding an extra variable and by solving the model in the quasi-static regime, see \aref{app:FEMmodel} for details.

\begin{figure}[t!hbp]
\centering
  \centering
  \includegraphics[width=\linewidth]{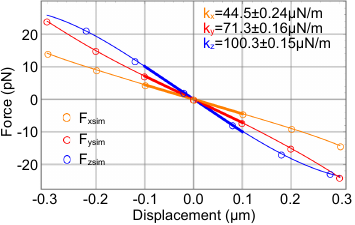}
  \caption{\xtxt{Note: we made the plot as wide as the column.} Force acting on a $R=0.5\,\muup$m sphere in the \AHC{}-trap from \fref{fig:AHCtrap}(c) as a function of the sphere's displacement relative to trap center. Open circles show the FEM-3D results, the thin solid line is a guide to the eye and the thick solid line shows a linear fit, from which we determine the spring constant $k_i$ and its uncertainty.}
  \label{fig:translationalstability}
\end{figure}

\begin{figure}[t!bhp]
\centering
  \includegraphics[width=\linewidth]{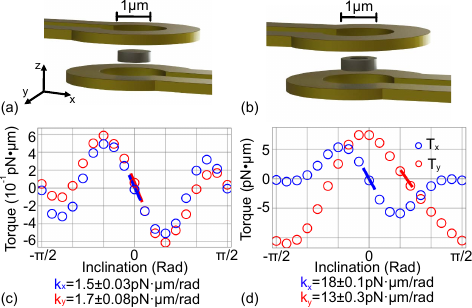}
\caption{Geometry of (a) cylinder and (b) ring in a realistic \AHC{}-trap. (c,d)  FEM-3D results of torque versus particle tilt. The x (y) component of the torque with respect to the y (x) axis is shown in blue (red). Zero tilt is defined by the particle orientation shown in (a,b). The corresponding angular frequencies are (c) $\omega_x^{\tau}=(14.0\pm 0.15)\,$kHz, $\omega_y^{\tau}=(15.0\pm 0.35)\,$kHz, and (d) $\omega_x^{\tau}=(60.0\pm 0.17)\,$kHz, $\omega_y^{\tau}=(52.1\pm 0.52)\,$kHz. Parameters are as in \fref{fig:AHCtrap} for the trap and \tref{tab:AHCfreqNonSphere} for the particles.}
\label{fig:angstability}
\end{figure}

\subsection{Anti-Helmholtz coil-trap}\label{sec:ahc}

We first analyze the magnetic field distribution of the \AHC{} trap. \Fref{fig:AHCtrap} shows the distributions obtained via the analytical formula for an empty \AHC{} [\fref{fig:AHCtrap}(d,g)], via FEM-2D [\fref{fig:AHCtrap}(e,h)] and via FEM-3D [\fref{fig:AHCtrap}(f,i)] for an \AHC{} with a superconducting sphere. As expected, the field distributions depend on the modeling used and, thus, will affect the trap frequency and levitation point.

\paragraph{Trap stability for translational \DOFF{}}

\begin{table}[t!bhp]
    \centering
    \begin{tabular}{cccc}
    \hline\hline\\[-5pt]
            & \multicolumn{3}{c}{Sphere $(\si{k\hertz})$} \\[1.5pt]
         Method   & $\omega_{x}/2\pi$  & $\omega_{y}/2\pi$  & $\omega_{z}/2\pi$ \\[1.5pt]
         \hline\\[-5pt]
            
    Point particle  & $24.8$&$24.8$&$49.6$  \\[1.5pt]
    
    Perfect diamagnet  & $-$&$-$&$45.7$  \\[1.5pt]
    
    Superconducting particle & $18.2$&$18.2$&$36.5$  \\[1.5pt]

    FEM-2D [3D]  &  $[17.8]$&$[17.8]$&$28.6$  \\[1.5pt]
    
    FEM-3D   &  $15.8$&$19.9$&$23.8$  \\[1.5pt]
    \hline\hline
    \end{tabular}
    \caption{Trap frequencies of a 1$\,\muup$m diameter sphere in the \AHC{} trap from \fref{fig:AHCtrap}. Note, $\omega_{x}$ and $\omega_{y}$ for FEM-2D were simulated with FEM-3D and a symmetric trap. The uncertainty on $\omega$ is below 0.14$\%$ and 0.7$\%$ for FEM-2D and FEM-3D, respectively.}
    \label{tab:AHCfreqSphere}
\end{table}

\begin{table}[t!bhp]
    \centering
    \begin{tabular}{c ccc ccc}
    \hline\hline\\[-5pt]
                  & \multicolumn{3}{c}{Cylinder $(\si{k\hertz})$} &  \multicolumn{3}{c}{Ring $(\si{k\hertz})$} \\[1.5pt]
                
         Method   & $\omega_{x}/2\pi$  & $\omega_{y}/2\pi$  & $\omega_{z}/2\pi$ & $\omega_{x}/2\pi$  & $\omega_{y}/2\pi$  & $\omega_{z}/2\pi$ \\ [1.5pt]
         \hline\\[-5pt]
            
    FEM-2D [3D]  &   $[16.6]$&$[16.6]$&$47.5$  & $[26.4]$&$[26.4]$&$49.1$  \\[1.5pt]
    
    FEM-3D  &  $12.6$&$17.2$&$36.8$  &  $25.4$&$26.8$&$37.9$ \\[1.5pt]
    \hline\hline
    \end{tabular}
    \caption{Trap frequencies of a cylinder (1$\,\muup$m diameter, 300$\,$nm height) and a ring (300$\,$nm thickness, inner and outer diameters of 0.5$\,\muup$m and 1$\,\muup$m, respectively) in the \AHC{} trap from \fref{fig:AHCtrap}. Note, $\omega_{x}$ and $\omega_{y}$ for FEM-2D were simulated with FEM-3D and a symmetric trap. The uncertainty on $\omega$ for the (cylinder, ring) is below (0.13$\%$, 0.13$\%$) and (1.3$\%$, 0.5$\%$) for FEM-2D and FEM-3D, respectively.}
    \label{tab:AHCfreqNonSphere}
\end{table}

The force acting on the spherical particle can now be calculated from the field distributions. \Fref{fig:translationalstability} shows the force acting on a superconducting sphere close to the center of the realistic \AHC{}-trap. At the center of the trap, the force equals zero as the magnetic force balances the gravitational force. The negative gradient of the force corresponds to a restoring force pushing the particle back to the trap center for small displacements. Thus, this parameter set results in a stably levitated particle. The thick solid lines are linear fits within $\pm100\,$nm of the trap position from which the spring constants $k_i$, their uncertainties and trap frequencies $\omega_i$ are calculated.

\paragraph{Trap stability for angular \DOFF{}}

When a non-spherical particle, such as a cylinder or ring, is placed in the field of the realistic \AHC{}-trap, torques also act on the particle, see \fref{fig:angstability}. Equilibrium orientations are found when the torque is zero and its slope negative, whereby the orientation with the largest slope is the stable and all others are metastable orientations. For a cylinder, a stable and metastable orientation are found at a tilt angle of 0 and $\pi/2$ with respect to the y axis, respectively. For the orientation with respect to the x axis, the stable orientation is close to 0, with a slight shift in angle due to the coil openings.

For a ring with no trapped flux, a stable and metastable orientation are found at a tilt angle of 0 and $\pi/2$ with respect to the y-axis, respectively. However, for the other orientation, there is only one stable orientation close to $\pi/6$. This asymmetry is caused by the coil openings and flux quantization that generates an additional current in the ring. A torque acts to minimize this current, orienting the ring towards the coil openings, where the field is weaker. If the \AHC{}-trap had no openings, a stable and metastable orientation would appear at an angle of 0 and $\pi/2$ with respect to the y-axis, respectively.

\begin{figure}[t!bhp]
    \centering
  \includegraphics[width=\linewidth]{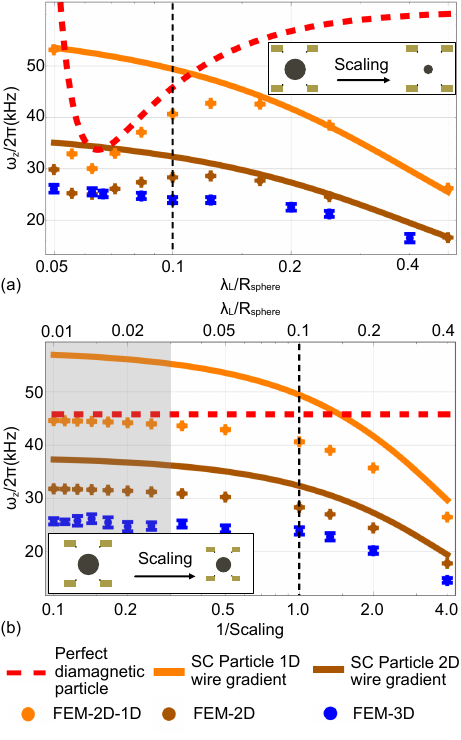}
   \caption{\xtxt{Note: we placed the legend below the plots.} Trap frequency of a superconducting sphere in a realistic \AHC{}. (a) The radius of the particle is scaled and the geometric parameters of the trap and $\lambda_L$ are kept constant, other parameters as in \tref{tab:AHCfreqSphere}. (b) The geometric parameters of the trap and particle are taken from \tref{tab:AHCfreqSphere} and scaled by a factor while the current density in the coils and $\lambda_L$ are kept constant. The vertical lines indicate the values for the initial geometry. The black points in the insets indicate the location of the 1D-current loops. \ntxt{The grey area represents geometries in which the particle is subject to magnetic fields above 80\,mT ($B_c$ of lead) with a maximal field of up to 230\,mT.} In \aref{app:addresults} we also consider the case when the 1D-current loops are centered in the wire.}
   \label{fig:AHCscaling}
\end{figure}

\paragraph{Trap frequency}

The previous analysis confirms that particles of different shapes can be stably levitated in a realistic \AHC{}. We now systematically study the trap frequency and consider first particles of different shape in the same \AHC{} trap, see \tref{tab:AHCfreqSphere} and \tref{tab:AHCfreqNonSphere}. We observe in \tref{tab:AHCfreqSphere} that the trap frequency for a spherical particle along z is by a factor of two larger than along x or y for the analytical models, which is expected due to the ideal anti Helmholtz coil arrangement in the trap. In FEM, however, this factor is reduced due to the deviation from a quadrupole field caused by the finite extent of the coil wires. We observe further that when accounting for the volume of the particle and treating it as a superconductor in the Meissner state, the magnetic field gradient around the particle is decreased and, thus, also the trap frequency. When also accounting for the opening of the coils via FEM-3D, the magnetic field distribution becomes asymmetric and leads to different trap frequencies along x and y.

\Tref{tab:AHCfreqNonSphere} shows that particles of non-spherical shape result in higher trap frequencies along the z axis. This difference can be attributed to the lower mass $m$ of the non-spherical particles as $\omega_t=\sqrt{k_t/m}$ (the diameter of all particles is the same). Additionally, the spring constant $k_t$ is also different due to the varying demagnetizing effect of each particle shape, for details see \aref{app:validate}.

We now analyze the dependence of the trap frequency on the size of a spherical particle in a trap with unaltered dimensions. In \fref{fig:AHCscaling}(a) we observe that for large particles the perfect diamagnetic sphere model yields similar results as FEM-2D-1D, since the normal conducting volume fraction of the particle is negligible compared to its superconducting volume fraction. Deviations occur when the particle radius is decreased to a size where magnetic field penetration becomes relevant, i.e., for $\lambda_L/R_{\mathrm{sphere}}\gtrapprox0.1$. When comparing FEM-2D-1D to a superconducting particle in a quadrupole field \cite{hofer_analytic_2019}, we observe that for small particle sizes FEM gives similar results. However, for larger particle sizes ($\lambda_L/R_{\mathrm{sphere}}\lessapprox 0.15$), the two methods give different results, which we attribute to the difference between a quadrupole field and the field generated by the wires, becoming more pronounced for larger particles \ntxt{(see also \fref{fig:2Dfield_comparison})}. When accounting for coils of finite extent via FEM-2D, the gradient of the field decreases compared to FEM-2D-1D and, thus, the trap frequency also decreases. Also in this case, assuming a superconducting sphere in a quadrupole field gives similar results for small particle sizes, but deviates for larger ones. When accounting for the opening of the trap wires via FEM-3D, the trap frequency further decreases, as expected.

\begin{figure}[t!bhp]
    \centering
    \includegraphics[width=\linewidth]{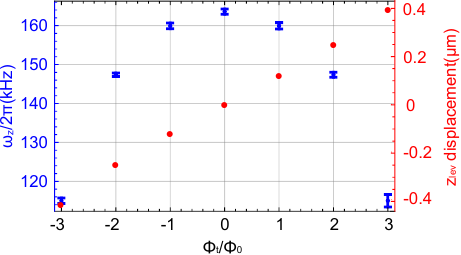}
    \caption{\xtxt{Note: we made the plot as wide as the column.} Levitated superconducting ring in an \AHC{}-trap simulated using FEM-2D. Trap frequency along z and levitation height with respect to the center of the trap as a function of trapped flux. Parameters as in \tref{tab:AHCfreqNonSphere}, but with an increased current in the coil wires to 100$\,$mA, i.e., $j_c=3.3\cdot10^{11}$A/m$^2$.}
    \label{fig:fluxquantdep}
\end{figure}

In \fref{fig:AHCscaling}(b) we analyze a scaled \AHC{}-trap architecture, whereby the dimensions of the particle and trap are simultaneously scaled, while keeping the current density in the coils and $\lambda_L$ constant. For large geometries, i.e., when the penetration depth is small compared to the particle size, the perfect diamagnetic particle method is in agreement with FEM-2D-1D. The decrease of the trap frequency for FEM-2D-1D when scaling down the system (for scaling factors $\lessapprox 3$, i.e., 1/scaling factor $\gtrapprox 0.3$) is due to the fact that for particles with a radius approaching $\lambda_L$ a portion of the sphere's volume becomes a normal conductor, and, thus, the magnetic force on the particle weakens. As before, when modeling the finite extent of the wires via FEM-2D the trap frequency decreases compared to FEM-2D-1D. For a superconducting sphere in a quadrupole field, we get similar results for small geometries, but deviations for large geometries. We attribute this behaviour as in \fref{fig:AHCscaling}(a) to the deviation of the field of the trap from a quadrupole field.

Levitation of a ring in the \AHC{} trap is particularly interesting. \Fref{fig:fluxquantdep} shows that the trap frequency and levitation height depend on the amount of trapped flux, $\Phi_t$, in the ring. The trap frequency decreases with increasing number of trapped flux, regardless of its orientation. The levitation height, however, increases monotonously with flux. This is because the ring seeks the region in the trap with a magnetic field strength that will generate the same flux as $\Phi_t$. As a result, the ring gets closer to one coil or the other depending on the orientation of $\Phi_t$, and, thus, further away from the trap center, where the field gradient is highest, reducing the trap frequency. 

To summarize, we find that FEM gives useful predictions for the stability, orientation and trap frequencies of different particle shapes levitated in realistic \AHC{} traps. In contrast, analytical models tend to overestimation of trap frequencies and deviating predictions when scaling the trap geometry, which can be traced back to the assumptions made by these models.

\begin{figure}[t!bhp]
\centering
  \includegraphics[width=\linewidth]{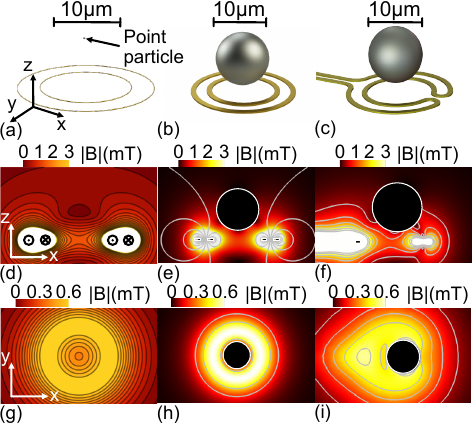}
\caption{Geometry of a \DLP{} trap for the (a)  point particle, (b) FEM-2D and (c) FEM-3D models and their corresponding field distributions for a cut in the XZ (d,e,f) and XY (g,h,i) plane that are tangent to the center of the trap. The white areas contain fields higher than the upper bound of the legend. Parameters:  inner coil diameter 12$\,\muup$m, outer coil diameter 18$\,\muup$m, applied current 38$\,$mA, 10$\,\muup$m diameter superconducting sphere with $\lambda_{\textrm{L}}=50\,$nm and $\rho=8570\,$kg/m$^3$. From FEM-2D we obtain trap gradients in the trap center of an empty trap of (65, 65, 125)$\,$\si{T/m} along the (x, y, z) axes.}
\label{fig:DLschematic}
\end{figure}

\subsection{\Dlp{} trap}\label{sec:dltrap}

We now turn to analyze the properties of the \DLP{} trap and show in \fref{fig:DLschematic} its magnetic field distribution. In \fref{fig:DLschematic}(d,g) the trap region is visible as the region surrounded by high field intensity. As can be seen in \fref{fig:DLschematic}(e,f,h,i) a particle with a diameter similar to the trap size fills up the trap region and is stable in the z direction due to gravity, since there is no magnetic field from above pushing it down. For these particle sizes, the \DLP{} trap is magneto-gravitational \cite{slezak_cooling_2018}. Hence, the simple layout of the \DLP{} trap comes at the expense of sacrificing magnetic field gradient and intensity.

\begin{table}[t!hbp]
    \centering
    \begin{tabular}{cccc}
     \hline\hline\\[-5pt]
    & \multicolumn{3}{c}{Sphere $(\si{\hertz})$} \\[1.5pt]
     Method       &  $\omega_x/2\pi$&$\omega_y/2\pi$&$\omega_z/2\pi$\\[1.5pt]
     \hline\\[-5pt]
    Point particle & 149.8 & 149.8 & 524.4 \\[1.5pt]
    Perfect diamagnet  & $-$ & $-$ & 465.1 \\[1.5pt]
    FEM-2D [3D]   & [113.3] & [113.3] & 423.2 \\[1.5pt]
    FEM-3D   & 82.3 & 119.8 & 355.0 \\[1.5pt]
    \hline\hline
    \end{tabular}
    \caption{Trap frequencies for a 10\,$\muup$m sphere levitated in a \dlp{} trap. Parameters as in \fref{fig:DLschematic}. Note, $\omega_{x}$ and $\omega_{y}$ for FEM-2D were simulated with FEM-3D and a symmetric trap. The uncertainty on $\omega$ is below 0.7$\%$ and 8.7$\%$ for FEM-2D and FEM-3D, respectively.}
    \label{tab:DLfreq}
\end{table}

The breaking of symmetry due to the openings of the coil wires has a significant effect in the \DLP{} trap. As shown in \fref{fig:DLschematic}(i), the field on the side of the current feed lines interferes constructively with the field generated by the inner coil, creating a higher field intensity at the left side of the particle that pushes it towards the direction of positive x. At the same time, the field opening at the opposite side weakens the field, creating a lower field intensity at the right side of the particle, which weakens the push in the direction of negative x towards the coil center. This effect can lead to the particle not being trapped. Thus, a careful design of the \DLP{} trap is required in order to achieve stable levitation. As a rule of thumb, the opening left between the wires should be smaller than the wire width of the coil.

\begin{figure}[b!hp]
    \centering
  \includegraphics[width=\linewidth]{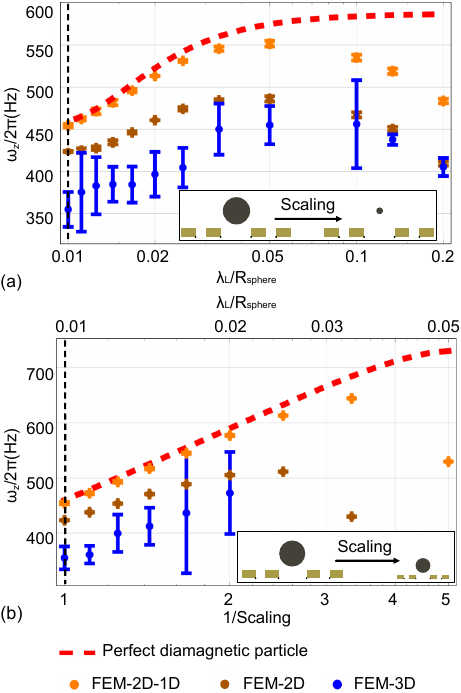}
   \caption{\xtxt{Note: we placed the legend below the plots.} Trap frequency of a superconducting sphere in a  \DLP{} trap. (a) The radius of the particle is scaled and the geometric parameters of the trap and $\lambda_L$ are kept constant, other parameters as in \fref{fig:DLschematic}. (b) The geometric parameters of the trap and particle are taken from \fref{fig:DLschematic} and scaled by a factor while the current density in the coils and $\lambda_L$ are kept constant. The vertical lines indicate the values for the initial geometry. The black points in the insets indicate the location of the 1D-current loops. Note that FEM-3D data points are only shown for trap geometries that result in stable levitation. In \aref{app:addresults} we also consider the case when the 1D-current loops are centered in the wire.}
   \label{fig:DLscaling}
\end{figure}

\paragraph{Trap frequency} 

\Tref{tab:DLfreq} shows trap frequencies for a 10$\,\muup$m spherical particle in a \DLP{} trap. The frequencies are below 1\,kHz and, thus, lower compared to the \AHC{} trap due to the \DLP{} trap being magneto-gravitational for this particle size. Note, the trap frequency will not change considerably for particles of a different shape, since any increase of the field gradient around the particle will push it higher up into regions of smaller magnetic field and, thus, smaller trap frequency.

\Fref{fig:DLscaling}(a) shows the trap frequency in the \DLP{} trap when changing particle size. For large particles, FEM-2D-1D agrees with the perfect diamagnetic particle method, while it deviates for smaller particles due to the finite field penetration. Modeling via FEM-2D and FEM-3D results in gradually smaller trap frequencies due to a reduced gradient of the trap. Interestingly, the trap frequency reaches a local maximum around $\lambda_L/R_{\mathrm{sphere}}\sim0.05$. For larger particles, the trap frequency decreases due to the trap becoming more magneto-gravitational, whereas for smaller particle sizes the magnetic field penetration into the particle leads to a reduction of the trap frequency.

In \fref{fig:DLscaling}(b) we consider a scaled system, where both the trap and the particle change size while keeping the current density of the trap and $\lambda_L$ constant. Again, we find agreement between the perfect diamagnetic particle method and the FEM-2D-1D for large geometries and an increasing discrepancy for smaller geometries due to magnetic field penetration. The trap frequency decreases in FEM-2D compared to FEM-2D-1D due to reducing the field gradient and further decreases when modeling via FEM-3D due to accounting for the wire opening. 

To summarize the analysis of the \DLP{} trap, we find that analytical models overestimate the trap frequency and may even fail to predict stability in case the wire coils have openings.

\section{Numerical analysis of flux-based read-out of particle motion}\label{sec:detect}

Magnetic levitation of superconducting micrometer-sized objects promises to reach an exceptional decoupling of the levitated object from its environment \cite{SClevitation,ringlevitation}. To verify this decoupling, one needs to detect the motion of the levitated particle. Motion detection can rely on flux-based read-out via a pick-up coil placed in the vicinity of the trap \cite{SCmagnetlevitation,vinante_ultralow_2020}. Particle oscillations around the trap center generate perturbations in the magnetic field distribution, which translate into a change of the magnetic flux threading through a pick-up coil. The pick-up coil could, in turn, be connected to a DC-SQUID, which converts the flux signal into a measurable voltage signal. The expected signal in a pick-up loop has been calculated analytically in previous work for the case of idealized situations \cite{SClevitation,SCmagnetlevitation}.

\begin{figure}[t!bhp]
\centering
\includegraphics[width=\linewidth]{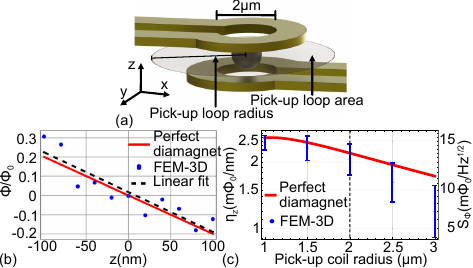}
\caption{Flux-based mechanical motion detection. (a) Geometry of the trap arrangement. The grey area represents the area over which the magnetic flux is integrated. (b) Flux threading through the grey area as a function of the particle displacement. Parameters as in \tref{tab:AHCfreqNonSphere}. (c) Signal strength $\eta_i$ and power spectral density $S_{\phi_{i}}$ detected by a pick-up loop of varying radius. The dashed line indicates the size of the pick-up loop as used in (b).}
\label{fig:detection}
\end{figure}

Using FEM we can now calculate the expected signal for realistic geometries by accounting for extended volumes, field penetration and flux quantization. In the following, we first consider a 1\,$\muup$m diameter spherical particle trapped in an \AHC{}-trap (cf.~\fref{fig:AHCtrap}). We are interested in calculating the magnetic flux threading a pick-up coil for small particle displacements with respect to the trap center, see \fref{fig:detection}(a). In \fref{fig:detection}(b) we compare the analytical prediction for a perfect diamagnetic sphere in a quadrupole field from Ref.~\cite{SClevitation} with our numerical FEM-3D results and find similar behaviour.

The slope of the curve in \fref{fig:detection}(b) yields the signal strength per displacement along direction $i$ (normalized by $10^{-3}\phi_0=1\,\mathrm{m}\phi_0$) as
\begin{equation}
   \eta_i=\frac{1}{\mathrm{m}\phi_0}\frac{\partial \phi}{\partial x_i}.
   \label{eq:signalstrength}
\end{equation}
Commonly, one measures the flux noise power spectral density $S_{\phi_i}(\omega)$, which is given as \cite{clerk_introduction_2010}:
\begin{equation}
S_{\phi_i}(\omega)=\eta_i S_{x_i}(\omega)=\eta_i x_{{\mathrm{rms}},i}\sqrt{\frac{\gamma_i}{(\omega-\omega_{i})^2+\gamma_i^2}},
\end{equation}
with $S_{x_i}(\omega)$ is the noise power spectral density of mechanical motion, $x_{{\mathrm{rms}},i}=\sqrt{k_{\textrm{B}} T/m\omega_{i}^2}$ ($k_{\textrm{B}}$ is Boltzmann's constant, $T$ is temperature) the root mean square amplitude of the oscillation in direction $i$ and $\gamma_i=\omega_i/Q_i$ is the mechanical damping with $Q_i$ being the mechanical quality factor. On mechanical resonance, one obtains $S_{\phi_i}(\omega_i)=\eta_ix_{{\mathrm{rms}},i}/\sqrt{\gamma_i}$.

\begin{table}[t!tbp]
    \centering
    \begin{tabular}{c ccc ccc ccc}
     \hline\hline\\[-5pt]
        &  $\eta_{x}$ &  $\eta_{y}$ & $\eta_{z}$  & $S_{\phi_{x}}$ & $S_{\phi_{y}}$ & $S_{\phi_{z}}$ & $S_{\phi_{0x}}$ & $S_{\phi_{0y}}$ & $S_{\phi_{0z}}$\\[1.5pt]
        &\multicolumn{3}{c}{$\left(10^{-1}{\mathrm{m}\phi_0}/{\mathrm{nm}}\right)$} & \multicolumn{3}{c}{$\left(\mathrm{m}{\phi_0}/{\sqrt{\mathrm{Hz}}}\right)$}
        & \multicolumn{3}{c}{$\left({\muup\phi_0}/{\sqrt{\mathrm{Hz}}}\right)$}\\[1.5pt]
        \hline\\[-5pt]
        Sphere   & 5.4 & 6.4 & 20.6 & 5.8 & 5.1 & 12.5 & 1.8 & 1.7 & 4.7 \\[1.5pt]
        Cylinder & 9.0 & 1.1 & 7.1 & 20.6 & 1.6 & 3.2 & 5.7 & 0.5 & 1.5\\[1.5pt]
        Ring     & 8.9 & 0.05 & 10.1 & 8.9 & 0.05 & 4.9 & 3.4 & 0.02 & 2.4\\[1.5pt]
        \hline\hline
    \end{tabular}
    \caption{Signal strength $\eta_i$ and noise power spectral density $S_{\phi_{i}}$ on mechanical resonance detected by a pick-up coil with 2$\,\muup$m  radius located between the two coils of the \AHC{}-trap. The dimensions of the trap and area of the pick-up coil are shown in \fref{fig:detection}(a). The trap and particle parameters are the same as in \tref{tab:AHCfreqNonSphere}. $S_{\phi_{x,y,z}}$ ($S_{\phi_{0x,0y,0z}}$) denotes the signal assuming $Q=10^7$ and $T=4$\,K (quantum ground state). The uncertainties are below 25$\%$ for the z direction and around 50$\%$ for the x and y directions.}
    \label{tab:AHCsignal}
\end{table}

\Tref{tab:AHCsignal} shows $\eta_i$ and $S_{\phi_i}(\omega_i)$ for a sphere, cylinder and ring in an \AHC{}-trap at a temperature of $4\,$K and for a conservative \cite{SClevitation,ringlevitation} $Q=10^7$. We also consider the case of detecting the ground state motion, i.e., $x_{0,i}=\sqrt{\hbar/m\omega_i}$, via measurement of flux, $S_{\phi_{0i}}(\omega_i)=\eta_ix_{0,i}/\sqrt{\gamma_i}$ ($\hbar$ is the reduced Planck's constant). The values are on the order of $\mathrm{m}\phi_0/\sqrt{\mathrm{Hz}}$ for thermally driven motion and some $\muup\phi_0/\sqrt{\mathrm{Hz}}$ for ground state motion. The former signals are well above the noise floor of state-of-the-art SQUID sensors, which are below $1\,\muup\phi_0/\sqrt{\textnormal{Hz}}$ for detection frequencies above 1\,kHz \cite{clarke_squid_2006,schurig_making_2014,wolbing_nb_2013}. While detection of ground state motion seems feasible, a further decrease in mechanical damping would be beneficial, as is predicted by theory \cite{SClevitation,SCmagnetlevitation}.

\Fref{fig:detection}(c) shows the signal strength and noise power spectral density when varying the pick-up coil radius. For small radii, the FEM results correspond within their uncertainty to the values predicted by Ref.~\cite{SClevitation}, but deviate for larger radii. This is because as the radius of the pick-up loop grows, the FEM model integrates over more coarsely meshed regions of the model and numerical errors accumulate.

\section{Conclusions}\label{sec:conclude}

We have analyzed in detail using analytical \cite{currentloop, lin_theoretical_2006,hofer_analytic_2019} and FEM modeling two promising trap architectures for levitating micrometer-sized superconducting particles in the Meissner state.  
The FEM modeling that we used is based on the $\vecc{A}$-V formulation \cite{cordier_3-d_1999,cordier_finite-element_1999,grilli_finite-element_2005,campbell_introduction_2011} and is generically applicable for superconductors in the Meissner state, such as for designing superconducting magnetic shields \cite{caputo_screening_2013} or filling factors in superconducting resonators \cite{niepce_geometric_2020}.

Crucially, we have shown that trap properties, like trap stability and frequency, can significantly differ from idealized, analytical models due to breaking of symmetry by coil openings, demagnetizing effects and flux quantization. We found that a chip-based \AHC{} trap is capable of levitating micrometer-sized particles of spherical, cylindrical and ring shape with trap frequencies well above 10\,kHz for a current density of $10^{11}\,$A/m$^2$ in the trap wires. However, the fabrication of such a trap on a single chip is complex and requires a three-layer process. A promising alternative would be to use a flip-chip architecture \cite{rosenberg_3d_2017}. In contrast, the \DLP{} trap is straight forward to fabricate in a single layer process. However, it comes at the expense of considerably lower trap frequencies of below 1\,kHz. Further, we confirmed numerically that read-out of the motion of the levitated particle using a pick-up loop in its vicinity \cite{SClevitation,SCmagnetlevitation} should lead to clearly detectable signals using presently available SQUID technology\cite{clarke_squid_2006,schurig_making_2014,wolbing_nb_2013}. We, thus, conclude that the analyzed chip-based superconducting traps are a viable approach for future quantum experiments that aim at levitating superconducting particles in the Meissner state \cite{SClevitation,ringlevitation,pino_-chip_2018}.

Extending our modeling by including flux pinning \cite{grilli_development_2013,morandi_5th_2017,grilli_dynamic_2018} via, for example, the critical state model \cite{bean_magnetization_1964,navau_macroscopic_2013} would allow studying alternative trap opportunities, which may offer chip-based traps with even higher trap frequencies. 

\begin{acknowledgments}

We acknowledge fruitful discussions with Jordi Prat Camps, Prasanna Venkatesh and COMSOL support. We thank Carles Navau and \`{A}lvar Sanchez for a critical reading of the manuscript. We are thankful for initial support in microfabrication from David Niepce. The work was supported by {Chalmers’ Excellence Initiative Nano} and the {Knut and Alice Wallenberg Foundation} through the Wallenberg Center for Quantum Technology (WACQT) and the Wallenberg Academy Fellow. Sample fabrication was performed in the Myfab Nanofabrication Laboratory at Chalmers. Simulations were performed on resources provided by the Swedish National Infrastructure for Computing (SNIC) at C3SE, Chalmers, partially funded by the Swedish Research Council through grant agreement no. 2016-07213.
\end{acknowledgments}

\appendix

\section{Magnetic levitation, forces and torques}
\label{app:maglev}

The goal of the chip-based traps is to stably levitate a superconducting particle in a point $\vecc{r}_{\mathrm{lev}}$ in free space above the surface of the chip. To this end, a local energy minimum in the potential energy landscape $U(\vecc{r})$ of the superconducting particle is required, with $U(\vecc{r})$ given by \cite{frog}:

\begin{equation}
    U(\vecc{r})=-\frac{1}{2}\int_V\vecc{M}(\vecc{r})\cdot\vecc{B}(\vecc{r})\,dV+mgz,
    \label{eq:potentialenergy}
\end{equation}
where $\vecc{M}$ is the magnetization, $\vecc{B}$ the magnetic field, $m$ the mass of the particle, $g$ the gravitational acceleration and $z$ is the height above the chip surface. The integration goes over the volume of the levitated particle. For illustration, let us assume the superconducting particle to be a perfect diamagnetic point particle with magnetic moment $\vecc{m}=V\,\vecc{M}=-V\,\vecc{B}/\muup_0$. Then, assuming $\vecc{B(\vecc{r})}$ depends linearly on $\vecc{r}$, the force acting on the particle is \cite{frog}:

\begin{equation}
    \vecc{F}(\vecc{r})=-\nabla U(\vecc{r})\approx-\frac{V}{\muup_0}\vecc{B}\cdot\nabla\vecc{B}-\rho Vg\vecc{\hat k}
    \label{eq:force}
\end{equation}

where $\vecc{\hat k}$ is the unit vector in the z direction, and we see that levitation is achieved when $\vecc{F}(\vecc{r}_{\mathrm{lev}})=0$, that is, when $\vecc{B}\cdot\nabla\vecc{B}=-\muup_0 g \rho\vecc{\hat k}$ at $\vecc{r}_{\mathrm{lev}}$. In reality, we cannot make the above approximation and we need to evaluate \eref{eq:potentialenergy} for an extended volume.
 
To this end, in our FEM model, the electromagnetic force and the torque on an object are calculated via the Maxwell stress tensor $T$, whose components $T_{ij}$ are given as:

\begin{eqnarray}
\nonumber T_{ij}&=&\epsilon_0\left(E_i E_j - \frac{1}{2}\delta_{ij}\lvert{\vecc{E}}\rvert^2 \right)\\
&&  +\frac{1}{\muup_0}\left( B_i B_j -\frac{1}{2}\delta_{ij}\lvert{\vecc{B}}\rvert^2 \right),
    \label{eq:maxstresstensor}
\end{eqnarray}

where $\epsilon_0$ and $\muup_0$ are the electrical permittivity and magnetic permeability, respectively, $E_i$ and $B_i$ are the vector components of the electric and the magnetic field and $\delta_{ij}$ is the Kronecker delta. The knowledge of the field distributions $\vecc{E}(\vecc{r})$ and $\vecc{B}(\vecc{r})$ is sufficient to calculate electromagnetic forces and torques via surface integrals as \cite{kovetz1990principles}

\begin{equation}
\vecc{F}=\oint_\Omega\vecc{n} T dS,
    \label{eq:magforcecalc}
\end{equation}
and
\begin{equation}
    \vecc{\tau}=\oint_\Omega  (\vecc{r}-\vecc{r}_0)\times(\vecc{n}T) dS,
    \label{eq:torquecalc}
\end{equation}

where $\vecc{\tau}$ is the torque, $\vecc{n}$ is the unit vector normal to the particle surface, $\Omega$ is the surface of the particle and $\vecc{r}$ and $\vecc{r}_0$ are the application point of the torque and the center of mass of the particle, respectively.

While balance of the gravitational and magnetic force is a necessary condition, it is not sufficient. Additionally, the local energy minimum at $\vecc{r}=(x,y,z)^T=\vecc{r}_{\mathrm{lev}}=(x_{\mathrm{lev}},y_{\mathrm{lev}},z_{\mathrm{lev}})^T$ must fulfill \cite{frog} $\partial^2 U(\vecc{r})/\partial x^2>0$, $\partial^2 U(\vecc{r})/\partial y^2>0$ and $\partial^2 U(\vecc{r})/\partial z^2>0$ in order to achieve stable levitation, so that the particle experiences a restoring force in the trap.

\section{FEM Modeling}
\label{app:FEMmodel}

The FEM simulations we use are based on the London model \cite{Tinkham} where, for small applied fields, the equation for the supercurrent in a superconductor can be written as\cite{anett_superconductivity}
\begin{equation}
    \vecc{J}_s=-\frac{1}{\muup_0\lambda_L^2}\vecc{A},
    \label{eq:jsl}
\end{equation}
where $\lambda_L=\sqrt{\frac{m}{\muup_0\lvert{\Psi}\rvert^2e^2}}$ is the London penetration depth, $\lvert{\Psi}\rvert^2=n_c$ is the squared amplitude of the order parameter's wave function $\Psi(\vecc{r},\theta)=\lvert\Psi(\vecc{r})\rvert e^{i\theta\vecc{r}}$ with phase $\theta$, $n_c$ is the Cooper pair density, and $e$ is the electron charge. By implementing this equation in FEM software as an external contribution to the current density in the superconductor domains, one can model domains as superconductors in the Meissner state. Note that \eref{eq:jsl} is in general not gauge invariant under the transformation $\vecc{A'}=\vecc{A}+\nabla\Phi_s$, where $\Phi_s$ is here an arbitrary scalar potential. However, in the specific case we consider, charge is conserved and the potentials $\vecc{A}$ and $\Phi_s$ change slowly in time (i.e., in the quasi-static regime), such that we can use \eref{eq:jsl} in the Coulomb gauge $\nabla\cdot\vecc{A}=0$.

The FEM implementation solves the Maxwell-London equations using $\vecc{A}$-V formulation \cite{cordier_3-d_1999,cordier_finite-element_1999,grilli_finite-element_2005,campbell_introduction_2011}. That is, the field equations are solved using the magnetic vector potential $\vecc{A}$ and the voltage V as the dependent variables. In our case, the field equations are solved in the quasi-static regime, so time derivatives of the equations describing the system are not involved. We would like to point out that describing dynamic systems is, however, possible as shown in Ref.~\cite{niepce_geometric_2020}. We note that, if $\vecc{B}$ is larger than the first critical field, $\vecc{B}_{c1}$, magnetic flux vortices will start nucleating in the superconductor. Thus $\vecc{B}_{c1}$ puts a bound on the maximal trap strength that can be studied in our modeling. 

Another feature of superconductivity is fluxoid quantization, which should be accounted for to accurately describe superconducting objects with holes. In our case, this concerns the levitation of ring-like particles. Fluxoid quantization can be derived by integrating the Ginzburg-Landau equation for the supercurrent\cite{Tinkham}
\begin{equation}
    \vecc{J}_s=i\frac{e\hbar}{2m_e}\left( \Psi^*\vecc{\nabla}\Psi-\Psi\vecc{\nabla}\Psi^* \right)-\frac{2e^2}{m_e}\vecc{A}\lvert{\Psi}\rvert^2,
   \label{eq:js}
\end{equation}
($m_e$ is the mass of the electron, $\hbar$ is Planck's constant) over a closed loop in the superconductor, which contains a hole with magnetic flux $\Phi_{\mathrm{hole}}$. This results in \cite{Tinkham}
\begin{eqnarray}
\nonumber \frac{m_e}{2e^2\lvert{\Psi}\rvert^2}\oint_C \vecc{J}_s\cdot d\vecc{l}&=&\frac{\hbar}{2e}\oint_C\vecc{\nabla}\theta \cdot d\vecc{l}-\oint_C\vecc{A}\cdot d\vecc{l}\\
&=&\Phi_0 n-\Phi_{\mathrm{hole}},
    \label{eq:fq}
\end{eqnarray}
where $n$ is an integer, and $\Phi_0=h/2e$ is the magnetic flux quantum. \Eref{eq:fq} tells us that the supercurrent will preserve the magnetic flux threading the hole of the superconductor as the multiple of $\Phi_0$ closest to $\Phi_{\mathrm{hole}}$.

\begin{figure}[t!bhp]
    \centering
    \includegraphics[width=\linewidth,keepaspectratio]{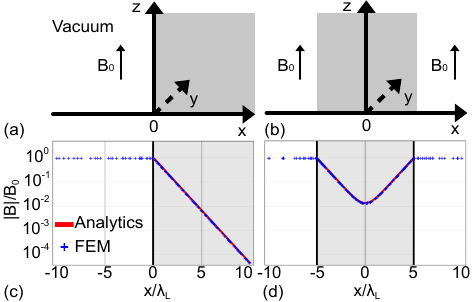}
    \caption{Magnetic field expulsion predicted by analytical equations (red line) and FEM (blue crosses) for (a) a semi-infinite superconductor and (b) a $t=1\,\muup$m thin film superconductor. The grey shaded areas represent the superconducting domain. The FEM data points are unevenly spaced due to the mesh of the model being finer at the vacuum-superconductor interface. Parameters used are: $\lambda_{\textrm{L}}=100\,$nm, $\vecc{B}_0=B_0\cdot\vecc{\hat k}$ with $B_0=100\,$mT.}
    \label{fig:Bdecay}
\end{figure}

Since our model does not account for the contributions of the wave function's gradient in \eref{eq:js}, fluxoid quantization cannot emerge from the implementation of \eref{eq:jsl}. We simplify our modeling by considering only flux quantization and, thus, neglect the flux in the ring's interior material caused by the finite penetration depth of the external magnetic field. This approximation is reasonable for  $\Lambda/R \ll 1$ (we have $\Lambda/R<0.04$), where $\Lambda=\lambda_L^2/d$ is the two dimensional effective penetration depth, $R$ is the lateral size of the superconducting object and $d$ its thickness \cite{SCrings}. We implement flux quantization ad hoc by defining the area of the hole in the superconductor over which \eref{eq:fq} is integrated, and impose an additional contribution to the current density of the superconductor such that the constraint
\begin{equation}
    \Phi_0 n-\Phi_{\mathrm{hole}}=0
    \label{eq:ge}
\end{equation}
is fulfilled within the defined area. In this way, a superconductor with trapped flux in a hole can be modeled. 

\section{Validation of FEM modeling}
\label{app:validate}

In order to validate our specific FEM implementation, we compare its results to test case, where analytical results exist. To this end, we select the magnetic field expulsion of a superconductor and demagnetizing effects of superconducting objects with different geometries. We also look at flux quantization in a ring and calculate the torque acting on a ring in a homogeneous magnetic field.

\paragraph{Magnetic Field Expulsion}

\begin{figure}[t!bhp]
\centering
  \includegraphics[width=0.5\textwidth]{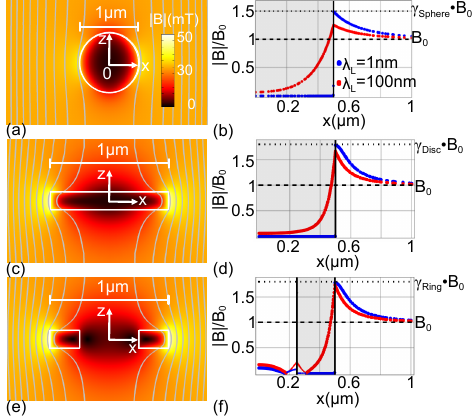}
\caption{Magnetic field distributions around a superconducting (a) sphere, (c) cylinder and (e) ring  under an external magnetic field $\vecc{B}_0=B_0\cdot\vecc{\hat k}$ with $B_0=30\,$mT. Panels (b,d,f) show the magnetic field intensity along the x-axis for a nearly perfect diamagnet ($\lambda_L\approx0)$ and a superconductor ($\lambda_L=100$\,nm). The black dashed line indicates the modulus of the applied field $\lvert{\vecc{B}_0}\rvert$ and the dotted line indicates the value of the field modulus according to the demagnetizing factor for each shape, $\gamma_{\mathrm{shape}}\cdot\lvert{\vecc{B}_0}\rvert$. The thinner lines in (f) indicate negative values. Parameters used: sphere of $1\,\muup$m diameter, cylinder of $1\,\muup$m diameter and 300\,nm thickness, ring of $1\,\muup$m outer diameter, $0.5\,\muup$m inner diameter and 300\,nm thickness}
\label{fig:demag}
\end{figure}

To examine magnetic field expulsion we consider (i) a flat \SC{} object with infinite extension in the z and positive x axes and (ii) a thin \SC{} film with infinite extension in the z axis, under a homogeneous magnetic field $\vecc{B}_0=B_0\cdot\vecc{\hat k}$, see \fref{fig:Bdecay}. For the first case, the Maxwell-London equations predict that $\vecc{B}_0$ is expected to decay exponentially within the superconductor with the characteristic length scale $\lambda_L$ (for superconductors with sizes $\gg\lambda_L$) \cite{Tinkham} $\vecc{B}\left(x\right)=\left(0,0,B_0e^{-\frac{x}{\lambda_L}}\right)^T$, where $x$ is the distance from the superconductor's surface. For the second case, the magnetic field inside a \SC{} thin film of thickness $t$ is expected to also decay exponentially from both sides, but the tails of each exponential will overlap in the middle of the thin film, thus limiting the field expulsion \cite{Tinkham}
\begin{equation}
    \label{eqn:doubledecay}
    \vecc{B}\left(x \right)=\left(0,0,B_0\frac{\cosh{\left(\frac{x}{\lambda_L}\right)}}{\cosh{\left(\frac{t}{2\lambda_L}\right)}}\right)^T.
\end{equation}

We simulate the structures for case (i) with a semi-infinite superconductor that occupies the positive half space $x >0$ and all z, and for case (ii) with a superconducting thin film with $t = 1\,\muup$m in x direction centered at zero while $y = z = \infty$. In both cases, we use $\lambda_{\textrm{L}} = 100\,$nm and a homogeneous magnetic field $\vecc{B}_0=B_0\cdot\vecc{\hat k}$ with $B_0=100\,$mT applied parallel to the $z$ axis. The results are shown in \fref{fig:Bdecay}(c,d) and show excellent agreement between FEM modeling and analytical equations.

\begin{figure}[t!bhp]
\centering
  \includegraphics[width=\linewidth]{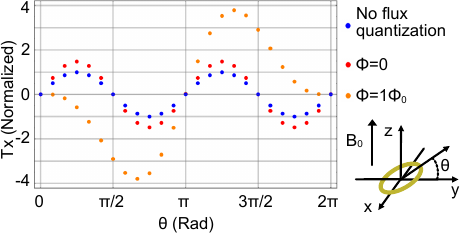}
  \caption{\xtxt{Note: we made the plot as wide as the column.} Normalized x component of the torque experienced by a superconducting ring (with inner and outer radii of 0.4 and 0.5$\,\muup$m, respectively, thickness of 50$\,$nm and $\lambda_L=50\,$nm) with (blue) no flux quantization, (red) with flux quantization and zero trapped flux, and (orange) with one trapped flux quantum parallel to the applied homogeneous field $\vecc{B}_0$. $\vecc{B}_0$ has a magnitude such that the flux through the ring equals $\Phi_0$ when $\theta=0$.}
  \label{fig:fluxquant}
\end{figure}

\paragraph{Demagnetizing Effects}

Field expulsion concentrates field lines around the surfaces of the superconducting object parallel to the field. In these regions, an increase of magnetic field intensity appears. This increase can be calculated analytically as a multiplying factor called demagnetizing factor. Demagnetizing effects arise naturally in our modeling. In \fref{fig:demag} we show the magnetic field distribution around a micrometer-sized sphere, cylinder and ring, under a homogeneous magnetic field $\vecc{B}_0=B_0\cdot\vecc{\hat k}$ with $B_0=30\,$mT. The demagnetizing factors for a perfect diamagnet with such geometries are 1.5, 1.8 and 1.8, respectively \cite{demag}. Our modeling as shown in \fref{fig:demag} perfectly matches the analytically calculated values when $\lambda_L$ is close to zero, i.e., for an ideal diamagnet. In the case of the ring, flux quantization is partly responsible for the magnetic field distribution within the ring. As indicated in \fref{fig:demag}(f), the thin section of the curve represents negative values of the magnetic field, which are generated by the supercurrent in the ring to keep $\Phi_{\mathrm{hole}}=0$.

\paragraph{Flux quantization: a ring in a homogeneous magnetic field.}

In general, generating a supercurrent has an energy cost. 
Then, it follows that the energy of the superconductor is minimized when the amount of supercurrent in it is smallest. Such an effect is shown in \fref{fig:fluxquant}, where we calculate the x component of the torque acting on a superconducting ring in a {\textit{homogeneous}} magnetic field $\vecc{B}_0$ as a function of the ring's inclination with respect to the y axis. We consider the cases for a superconducting ring with (i) no flux quantization, (ii) flux quantization with zero flux trapped and (iii) one flux quantum trapped with the same orientation as $\vecc{B}_0$.

The ring with no flux quantization experiences a torque because the field is less perturbed when $\vecc{B}_0$ is parallel to the area of the hole than when it is perpendicular. Hence, it takes less supercurrent to expel the field when $\theta=\pi/2$ or $3\pi/2$. When the area of the hole is perpendicular to $\vecc{B}_0$ the torque on the ring vanishes due to symmetry, since it is as likely to tilt clockwise or counter-clockwise, in other words, it is in an unstable equilibrium. The stable configuration for the ring including flux quantization and no trapped flux, i.e., $\Phi_t=0$, is to be oriented so that no flux is threading the hole, i.e., $\pi/2$ or $3\pi/2$. The difference is that the torque is stronger due to additional current from flux quantization that keeps $\Phi_t=0$ when $\theta \neq \pi/2$ or $3\pi/2$. For the case of a ring with one trapped flux quantum parallel to $\vecc{B}_0$, the configuration in which the least supercurrent is generated is that where $\vecc{B}_0$ is parallel to the trapped flux quantum, since $\vecc{B}_0$ is chosen so that the flux through the hole equals $\Phi_0$ when the ring is perpendicular to the field. Thus, the ring will experience a torque that will force it to $\theta=0$. For $\theta=\pi$ the ring will be unstable because the flux through the hole at this configuration is maximum ($\Phi=2\Phi_0$).

\begin{figure}[t!bhp]
\centering
\includegraphics[width=\linewidth]{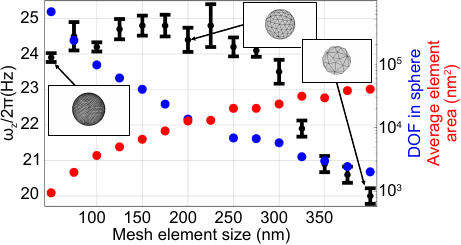}
\caption{\xtxt{Note: we made the plot as wide as the column.} Trap frequency of a spherical particle in an \AHC{} trap as considered in \fref{fig:AHCtrap} as a function of the mesh element size $l_{\mathrm{mesh}}$ on the particle's surface using FEM-3D. The insets show the mesh on the surface of the particle at the given element sizes.}
\label{fig:meshdependence}
\end{figure}

\begin{figure}[t!bhp]
    \centering
    \includegraphics[width=\linewidth]{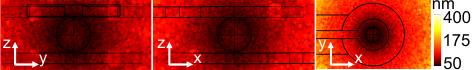}
    \caption{Mesh element size of the FEM-3D model for a superconducting sphere in an \AHC{}-trap, with the maximum element size on the particle surface set to 50$\,$nm. Parameters as in \fref{fig:AHCtrap}.}
    \label{fig:AHCmesh}
\end{figure}

\paragraph{Flux quantization: a ring and levitation}

Ref.~\cite{carles_2020} provides an analytical formula for the trap frequency along the vertical direction for levitating a ring in a quadrupole field, including flux quantization. We compared FEM-2D simulations to this formula for a ring with inner and outer radii of 0.4$\,\muup$m and 0.5$\,\muup$m, respectively, thickness of 50$\,$nm and $\lambda_L=50\,$nm in an \AHC{} trap with coil radius and separation of 10$\,\muup$m and a current of 3$\,$A. Using FEM-2D and assuming zero flux trapped in the ring, we obtain ($212\pm 0.6$)\,\si{kHz}, which is in good agreement with the 209$\,$\si{kHz} predicted by Ref.~\cite{carles_2020}. We also calculated the inductance of such a superconducting ring with flux quantization with FEM and obtained $(2\pm0.14)$\,\si{pH}, which is in good agreement with $1.6\,$\si{pH} predicted by Ref.~\cite{SCrings}. 

\section{FEM meshing}
\label{app:FEMmesh}

\begin{figure}[t!bhp]
\centering
\includegraphics[width=\linewidth]{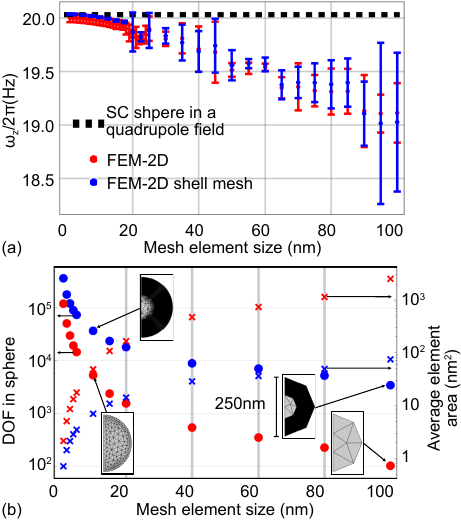}
\caption{\xtxt{Note: we made the plot as wide as the column.} Trap frequency of a spherical particle with 125$\,$nm radius and $\lambda_L=50\,$nm in an \AHC{} trap as considered in \fref{fig:AHCscaling} with a scaling factor of 10, as a function of the maximal mesh element size $l_{\mathrm{mesh}}$. Comparison of (a) FEM-2D to analytical results obtained for the configuration of a superconducting sphere in a quadrupole field \cite{hofer_analytic_2019} for two different meshing strategies: (i) triangular mesh only (red data) and (ii) triangular mesh combined with a shell mesh that meshes the outermost volume of 75\,nm thickness of the sphere with onion-type layers of 1\,nm thickness (blue data). (b) Degrees of freedom (dots) and average element area (crosses) in the sphere for each of the meshing strategies.}
\label{fig:meshdependence2Dsmall}
\end{figure}

\begin{figure}[t!bhp]
\centering
\includegraphics[width=\linewidth]{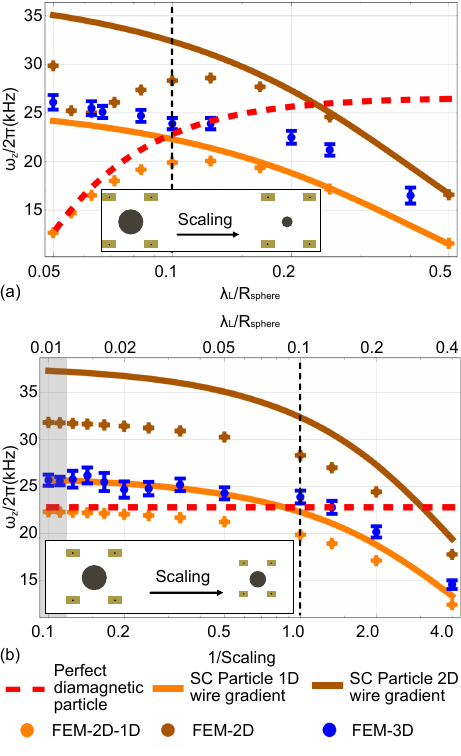}
\caption{\xtxt{Note: we placed the legend below the plots.}  Trap frequency of a superconducting sphere in a realistic \AHC{} predicted by the image method \cite{lin_theoretical_2006}, by assuming a superconducting sphere in a quadrupole field \cite{hofer_analytic_2019}, FEM-2D with 1D-wires, FEM-2D and FEM-3D. (a) The radius of the particle is scaled and the geometrical parameters of the trap and $\lambda_L$ are kept constant with parameters as given in \tref{tab:AHCfreqSphere}. (b) The geometrical parameters of the trap and particle are taken from \tref{tab:AHCfreqSphere} and scaled by a scaling factor while the current density in the coils and $\lambda_L$ are kept constant. The vertical lines indicate the initial geometry. The black points in the schematics indicate the location of the 1D-current loops. \ntxt{The grey area represents geometries in which the particle is subject to magnetic fields above 80\,mT ($B_c$ of lead) with a maximal field of up to 100\,mT.}}
\label{fig:AHCscalingcentered}
\end{figure}

Given that the model is based on FEM, the results are mesh dependent. Constructing a mesh fine enough at the surface of the superconducting domains is critical to get reliable results. This dependence is illustrated in \fref{fig:meshdependence}, where the trap frequency along z for a 1$\,\muup$m diameter sphere in an \AHC{} trap (cf.~\fref{fig:AHCtrap}) is calculated via FEM-3D. For these simulations we changed the maximal allowed mesh element size, $l_{\mathrm{mesh}}$, on the surface of the particle resulting in gradually finer meshed particles, see the insets in \fref{fig:meshdependence} and \fref{fig:AHCmesh}. When reducing $l_{\mathrm{mesh}}$, the FEM meshing algorithm gradually increases the number of mesh elements in the sphere and, thus, reduces the average element area that one mesh element covers. This is reflected in the number of degrees of freedom (DOF) in the sphere, that is, the number of unknowns to solve for in the model, which in general equals the number of dependent variables ($A_x$, $A_y$, $A_z$ and the gauge fixing potential inside the sphere) times the number of nodes in the geometry. In all our simulations we use quadratic mesh discretization, which means the lines connecting the mesh nodes are not straight lines but polynomials of second order. For $l_{\mathrm{mesh}}\lessapprox 5\cdot \lambda_L=250\,$nm, we observe no clear trend of the trap frequency within its uncertainty. However, for $l_{\mathrm{mesh}}\gtrapprox 0.5\cdot R_{\mathrm{sphere}}=250\,$nm, the particle itself is not properly resolved and the magnetic field penetrates parts or the entire volume of the particle, which effectively increases the effect of field penetration and, thus, decreases the trap frequency.

For FEM-2D we can decrease $l_{\mathrm{mesh}}$ further as the computational cost is not as large as for FEM-3D simulations. \Fref{fig:meshdependence2Dsmall} shows the trap frequency of a 150$\,$nm radius sphere in an \AHC{} trap in dependence of $l_{\mathrm{mesh}}$. For fine enough meshing, i.e., $l_{\mathrm{mesh}}\lessapprox10\,$nm corresponding to  $>10^4$ DOF, the FEM simulations converge to the analytical results obtained for a superconducting sphere in a quadrupole field. The small discrepancy is attributed to the difference between the field distribution of a  quadrupole field and the field of the modeled trap.

The trap frequency dependence on the mesh might not only be related to the mesh element size itself, but also on differences in the mesh being differently built for similar FEM models. To test this, we simulated the trap configuration as used for \fref{fig:meshdependence} for slightly different $l_{\mathrm{mesh}}$ of (49.9, 49.95, 50.00, 50.05, 50.1)$\,$nm and get trap frequencies of (23.7, 23.6, 23.8, 23.6, 23.6)\,kHz, resulting in a mean value of $(23.66\pm0.09)\,$kHz. Thus, the scatter of trap frequency of about $\pm0.4\%$ from using nearly similar meshes is smaller than the fit uncertainty of the trap frequency.

Note that the computation time for obtaining a typical magnetic field distribution of a particle is $30-120$\,min and requires $50-600$\,GB of RAM \ntxt{on computing nodes with 20 core Intel E5-2650v3 CPUs with 2.30\,GHz base frequency available via a computing cluster.}

\section{Additional FEM results}
\label{app:addresults}

\begin{figure}[t!bhp]
\centering
\includegraphics[width=\linewidth]{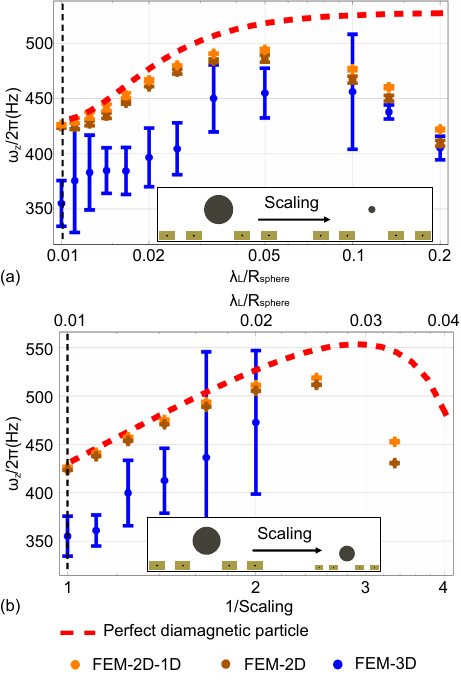}
\caption{\xtxt{Note: we placed the legend below the plots.}  Trap frequency dependence of a superconducting sphere trapped in a realistic \dlp{} trap. (a) The radius of the particle is scaled and the geometrical parameters of the trap are kept constant. (b) The geometrical parameters of the trap and the particle are taken from \fref{fig:DLschematic} and scaled by a scaling factor while the current density in the coils and $\lambda_L$ are kept constant. The vertical lines indicate the initial geometry. The black points in the schematics indicate the location of the 1D-current loops. Note that FEM-3D data points are only shown for trap geometries that result in stable levitation.}
\label{fig:DLscalingcentered}
\end{figure}

\subsection{Results for centered 1D current loops}

\Fref{fig:AHCscalingcentered} and \fref{fig:DLscalingcentered} show the dependence of the trap frequency with the scaling of the geometry of the respective trap. Here, we place the 1D current loops in FEM2D-1D and the perfect diamagnetic particle method at a position corresponding to the center of the wires. This data can be compared to the corresponding data shown in \fref{fig:AHCscaling} and \fref{fig:DLscaling} when the 1D current loops are placed at the innermost corner of the coils.

\subsection{Field distribution in \AHC{} trap}

\begin{figure}[b!htp]
    \centering
    \includegraphics[width=\linewidth]{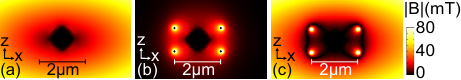}
    \caption{\xtxt{This figure is new.} \ntxt{Field distributions for an \AHC{} trap with a superconducting sphere calculated from (a) an ideal quadrupole field and (b) the field generated by infinitesimally small wires obtained via FEM-2D-1D. (c) shows the absolute difference between the two distributions. Parameters are: AHC coils with 1$\,\mu$m radius,  1$\,\mu$m coil separation, superconducting sphere of 0.5$\,\mu$m radius,  $\lambda_\mathrm{L}=$50$\,$nm, current through coil wires is 30\,mA.}}
    \label{fig:2Dfield_comparison}
\end{figure}

\ntxt{\Fref{fig:2Dfield_comparison} shows the magnetic field distribution in an \AHC{} trap for the case of a superconducting sphere in a quadruple field \cite{hofer_analytic_2019} and in the field generated by quasi-1D wires (obatined via FEM-2D-1D). These field distributions are similar close to the particle surface, but deviate much more when approaching the coil wires.}

\clearpage

\bibliography{refsfem}

\begin{thebibliography}{57}%
\makeatletter
\providecommand \@ifxundefined [1]{%
 \@ifx{#1\undefined}
}%
\providecommand \@ifnum [1]{%
 \ifnum #1\expandafter \@firstoftwo
 \else \expandafter \@secondoftwo
 \fi
}%
\providecommand \@ifx [1]{%
 \ifx #1\expandafter \@firstoftwo
 \else \expandafter \@secondoftwo
 \fi
}%
\providecommand \natexlab [1]{#1}%
\providecommand \enquote  [1]{``#1''}%
\providecommand \bibnamefont  [1]{#1}%
\providecommand \bibfnamefont [1]{#1}%
\providecommand \citenamefont [1]{#1}%
\providecommand \href@noop [0]{\@secondoftwo}%
\providecommand \href [0]{\begingroup \@sanitize@url \@href}%
\providecommand \@href[1]{\@@startlink{#1}\@@href}%
\providecommand \@@href[1]{\endgroup#1\@@endlink}%
\providecommand \@sanitize@url [0]{\catcode `\\12\catcode `\$12\catcode
  `\&12\catcode `\#12\catcode `\^12\catcode `\_12\catcode `\%12\relax}%
\providecommand \@@startlink[1]{}%
\providecommand \@@endlink[0]{}%
\providecommand \url  [0]{\begingroup\@sanitize@url \@url }%
\providecommand \@url [1]{\endgroup\@href {#1}{\urlprefix }}%
\providecommand \urlprefix  [0]{URL }%
\providecommand \Eprint [0]{\href }%
\providecommand \doibase [0]{http://dx.doi.org/}%
\providecommand \selectlanguage [0]{\@gobble}%
\providecommand \bibinfo  [0]{\@secondoftwo}%
\providecommand \bibfield  [0]{\@secondoftwo}%
\providecommand \translation [1]{[#1]}%
\providecommand \BibitemOpen [0]{}%
\providecommand \bibitemStop [0]{}%
\providecommand \bibitemNoStop [0]{.\EOS\space}%
\providecommand \EOS [0]{\spacefactor3000\relax}%
\providecommand \BibitemShut  [1]{\csname bibitem#1\endcsname}%
\let\auto@bib@innerbib\@empty
\bibitem [{\citenamefont {Moon}\ and\ \citenamefont
  {Chang}(1994)}]{moon_superconducting_1994}%
  \BibitemOpen
  \bibfield  {author} {\bibinfo {author} {\bibfnamefont {F.C.}\ \bibnamefont
  {Moon}}\ and\ \bibinfo {author} {\bibfnamefont {P.Z.}\ \bibnamefont
  {Chang}},\ }\href@noop {} {\emph {\bibinfo {title} {Superconducting
  {Levitation}: {Applications} to {Bearings} and {Magnetic}
  {Transportation}}}}\ (\bibinfo  {publisher} {Wiley},\ \bibinfo {year}
  {1994})\BibitemShut {NoStop}%
\bibitem [{\citenamefont {Brandt}(1989)}]{Brandt}%
  \BibitemOpen
  \bibfield  {author} {\bibinfo {author} {\bibfnamefont {E.~H.}\ \bibnamefont
  {Brandt}},\ }\bibfield  {title} {\enquote {\bibinfo {title} {Levitation in
  physics},}\ }\href {\doibase 10.1126/science.243.4889.349} {\bibfield
  {journal} {\bibinfo  {journal} {Science}\ }\textbf {\bibinfo {volume}
  {243}},\ \bibinfo {pages} {349--355} (\bibinfo {year} {1989})}\BibitemShut
  {NoStop}%
\bibitem [{\citenamefont {Arkadiev}(1947)}]{arkadiev_floating_1947}%
  \BibitemOpen
  \bibfield  {author} {\bibinfo {author} {\bibfnamefont {V.}~\bibnamefont
  {Arkadiev}},\ }\bibfield  {title} {\enquote {\bibinfo {title} {A {Floating}
  {Magnet}},}\ }\href {\doibase 10.1038/160330a0} {\bibfield  {journal}
  {\bibinfo  {journal} {Nature}\ }\textbf {\bibinfo {volume} {160}},\ \bibinfo
  {pages} {330} (\bibinfo {year} {1947})}\BibitemShut {NoStop}%
\bibitem [{\citenamefont {Goodkind}(1999)}]{goodkind_superconducting_1999}%
  \BibitemOpen
  \bibfield  {author} {\bibinfo {author} {\bibfnamefont {John~M.}\ \bibnamefont
  {Goodkind}},\ }\bibfield  {title} {\enquote {\bibinfo {title} {The
  superconducting gravimeter},}\ }\href {\doibase 10.1063/1.1150092} {\bibfield
   {journal} {\bibinfo  {journal} {Review of Scientific Instruments}\ }\textbf
  {\bibinfo {volume} {70}},\ \bibinfo {pages} {4131} (\bibinfo {year}
  {1999})}\BibitemShut {NoStop}%
\bibitem [{\citenamefont {Romero-Isart}\ \emph {et~al.}(2012)\citenamefont
  {Romero-Isart}, \citenamefont {Clemente}, \citenamefont {Navau},
  \citenamefont {Sanchez},\ and\ \citenamefont {Cirac}}]{SClevitation}%
  \BibitemOpen
  \bibfield  {author} {\bibinfo {author} {\bibfnamefont {O.}~\bibnamefont
  {Romero-Isart}}, \bibinfo {author} {\bibfnamefont {L.}~\bibnamefont
  {Clemente}}, \bibinfo {author} {\bibfnamefont {C.}~\bibnamefont {Navau}},
  \bibinfo {author} {\bibfnamefont {A.}~\bibnamefont {Sanchez}}, \ and\
  \bibinfo {author} {\bibfnamefont {J.~I.}\ \bibnamefont {Cirac}},\ }\bibfield
  {title} {\enquote {\bibinfo {title} {Quantum magnetomechanics with levitating
  superconducting microspheres},}\ }\href {\doibase
  10.1103/PhysRevLett.109.147205} {\bibfield  {journal} {\bibinfo  {journal}
  {Phys. Rev. Lett.}\ }\textbf {\bibinfo {volume} {109}},\ \bibinfo {pages}
  {147205} (\bibinfo {year} {2012})}\BibitemShut {NoStop}%
\bibitem [{\citenamefont {Cirio}\ \emph {et~al.}(2012)\citenamefont {Cirio},
  \citenamefont {Brennen},\ and\ \citenamefont {Twamley}}]{ringlevitation}%
  \BibitemOpen
  \bibfield  {author} {\bibinfo {author} {\bibfnamefont {M.}~\bibnamefont
  {Cirio}}, \bibinfo {author} {\bibfnamefont {G.~K.}\ \bibnamefont {Brennen}},
  \ and\ \bibinfo {author} {\bibfnamefont {J.}~\bibnamefont {Twamley}},\
  }\bibfield  {title} {\enquote {\bibinfo {title} {Quantum magnetomechanics:
  Ultrahigh-$q$-levitated mechanical oscillators},}\ }\href {\doibase
  10.1103/PhysRevLett.109.147206} {\bibfield  {journal} {\bibinfo  {journal}
  {Phys. Rev. Lett.}\ }\textbf {\bibinfo {volume} {109}},\ \bibinfo {pages}
  {147206} (\bibinfo {year} {2012})}\BibitemShut {NoStop}%
\bibitem [{\citenamefont {Johnsson}\ \emph {et~al.}(2016)\citenamefont
  {Johnsson}, \citenamefont {Brennen},\ and\ \citenamefont
  {Twamley}}]{johnsson_macroscopic_2016}%
  \BibitemOpen
  \bibfield  {author} {\bibinfo {author} {\bibfnamefont {Mattias~T.}\
  \bibnamefont {Johnsson}}, \bibinfo {author} {\bibfnamefont {Gavin~K.}\
  \bibnamefont {Brennen}}, \ and\ \bibinfo {author} {\bibfnamefont {Jason}\
  \bibnamefont {Twamley}},\ }\bibfield  {title} {\enquote {\bibinfo {title}
  {Macroscopic superpositions and gravimetry with quantum magnetomechanics},}\
  }\href {\doibase 10.1038/srep37495} {\bibfield  {journal} {\bibinfo
  {journal} {Scientific Reports}\ }\textbf {\bibinfo {volume} {6}},\ \bibinfo
  {pages} {37495} (\bibinfo {year} {2016})}\BibitemShut {NoStop}%
\bibitem [{\citenamefont {Pino}\ \emph {et~al.}(2018)\citenamefont {Pino},
  \citenamefont {Prat-Camps}, \citenamefont {Sinha}, \citenamefont
  {Venkatesh},\ and\ \citenamefont {Romero-Isart}}]{pino_-chip_2018}%
  \BibitemOpen
  \bibfield  {author} {\bibinfo {author} {\bibfnamefont {H.}~\bibnamefont
  {Pino}}, \bibinfo {author} {\bibfnamefont {J.}~\bibnamefont {Prat-Camps}},
  \bibinfo {author} {\bibfnamefont {K.}~\bibnamefont {Sinha}}, \bibinfo
  {author} {\bibfnamefont {B.~Prasanna}\ \bibnamefont {Venkatesh}}, \ and\
  \bibinfo {author} {\bibfnamefont {O.}~\bibnamefont {Romero-Isart}},\
  }\bibfield  {title} {\enquote {\bibinfo {title} {On-chip quantum interference
  of a superconducting microsphere},}\ }\href {\doibase
  10.1088/2058-9565/aa9d15} {\bibfield  {journal} {\bibinfo  {journal} {Quantum
  Sci. Technol.}\ }\textbf {\bibinfo {volume} {3}},\ \bibinfo {pages} {025001}
  (\bibinfo {year} {2018})}\BibitemShut {NoStop}%
\bibitem [{\citenamefont {Prat-Camps}\ \emph {et~al.}(2017)\citenamefont
  {Prat-Camps}, \citenamefont {Teo}, \citenamefont {Rusconi}, \citenamefont
  {Wieczorek},\ and\ \citenamefont {Romero-Isart}}]{SCmagnetlevitation}%
  \BibitemOpen
  \bibfield  {author} {\bibinfo {author} {\bibfnamefont {J.}~\bibnamefont
  {Prat-Camps}}, \bibinfo {author} {\bibfnamefont {C.}~\bibnamefont {Teo}},
  \bibinfo {author} {\bibfnamefont {C.~C.}\ \bibnamefont {Rusconi}}, \bibinfo
  {author} {\bibfnamefont {W.}~\bibnamefont {Wieczorek}}, \ and\ \bibinfo
  {author} {\bibfnamefont {O.}~\bibnamefont {Romero-Isart}},\ }\bibfield
  {title} {\enquote {\bibinfo {title} {Ultrasensitive inertial and force
  sensors with diamagnetically levitated magnets},}\ }\href {\doibase
  10.1103/PhysRevApplied.8.034002} {\bibfield  {journal} {\bibinfo  {journal}
  {Phys. Rev. Applied}\ }\textbf {\bibinfo {volume} {8}},\ \bibinfo {pages}
  {034002} (\bibinfo {year} {2017})}\BibitemShut {NoStop}%
\bibitem [{\citenamefont {Jackson~Kimball}\ \emph {et~al.}(2016)\citenamefont
  {Jackson~Kimball}, \citenamefont {Sushkov},\ and\ \citenamefont
  {Budker}}]{jackson_kimball_precessing_2016}%
  \BibitemOpen
  \bibfield  {author} {\bibinfo {author} {\bibfnamefont {Derek~F.}\
  \bibnamefont {Jackson~Kimball}}, \bibinfo {author} {\bibfnamefont
  {Alexander~O.}\ \bibnamefont {Sushkov}}, \ and\ \bibinfo {author}
  {\bibfnamefont {Dmitry}\ \bibnamefont {Budker}},\ }\bibfield  {title}
  {\enquote {\bibinfo {title} {Precessing {Ferromagnetic} {Needle}
  {Magnetometer}},}\ }\href {\doibase 10.1103/PhysRevLett.116.190801}
  {\bibfield  {journal} {\bibinfo  {journal} {Phys. Rev. Lett.}\ }\textbf
  {\bibinfo {volume} {116}},\ \bibinfo {pages} {190801} (\bibinfo {year}
  {2016})}\BibitemShut {NoStop}%
\bibitem [{\citenamefont {Slezak}\ \emph {et~al.}(2018)\citenamefont {Slezak},
  \citenamefont {Lewandowski}, \citenamefont {Hsu},\ and\ \citenamefont
  {D'Urso}}]{slezak_cooling_2018}%
  \BibitemOpen
  \bibfield  {author} {\bibinfo {author} {\bibfnamefont {Bradley~R.}\
  \bibnamefont {Slezak}}, \bibinfo {author} {\bibfnamefont {Charles~W.}\
  \bibnamefont {Lewandowski}}, \bibinfo {author} {\bibfnamefont {Jen-Feng}\
  \bibnamefont {Hsu}}, \ and\ \bibinfo {author} {\bibfnamefont {Brian}\
  \bibnamefont {D'Urso}},\ }\bibfield  {title} {\enquote {\bibinfo {title}
  {Cooling the motion of a silica microsphere in a magneto-gravitational trap
  in ultra-high vacuum},}\ }\href {\doibase 10.1088/1367-2630/aacac1}
  {\bibfield  {journal} {\bibinfo  {journal} {New J. Phys.}\ }\textbf {\bibinfo
  {volume} {20}},\ \bibinfo {pages} {063028} (\bibinfo {year}
  {2018})}\BibitemShut {NoStop}%
\bibitem [{\citenamefont {Wang}\ \emph {et~al.}(2019)\citenamefont {Wang},
  \citenamefont {Lourette}, \citenamefont {O’Kelley}, \citenamefont {Kayci},
  \citenamefont {Band}, \citenamefont {Kimball}, \citenamefont {Sushkov},\ and\
  \citenamefont {Budker}}]{wang_dynamics_2019}%
  \BibitemOpen
  \bibfield  {author} {\bibinfo {author} {\bibfnamefont {Tao}\ \bibnamefont
  {Wang}}, \bibinfo {author} {\bibfnamefont {Sean}\ \bibnamefont {Lourette}},
  \bibinfo {author} {\bibfnamefont {Sean~R.}\ \bibnamefont {O’Kelley}},
  \bibinfo {author} {\bibfnamefont {Metin}\ \bibnamefont {Kayci}}, \bibinfo
  {author} {\bibfnamefont {Y.B.}\ \bibnamefont {Band}}, \bibinfo {author}
  {\bibfnamefont {Derek F.~Jackson}\ \bibnamefont {Kimball}}, \bibinfo {author}
  {\bibfnamefont {Alexander~O.}\ \bibnamefont {Sushkov}}, \ and\ \bibinfo
  {author} {\bibfnamefont {Dmitry}\ \bibnamefont {Budker}},\ }\bibfield
  {title} {\enquote {\bibinfo {title} {Dynamics of a {Ferromagnetic} {Particle}
  {Levitated} over a {Superconductor}},}\ }\href {\doibase
  10.1103/PhysRevApplied.11.044041} {\bibfield  {journal} {\bibinfo  {journal}
  {Phys. Rev. Applied}\ }\textbf {\bibinfo {volume} {11}},\ \bibinfo {pages}
  {044041} (\bibinfo {year} {2019})}\BibitemShut {NoStop}%
\bibitem [{\citenamefont {Timberlake}\ \emph {et~al.}(2019)\citenamefont
  {Timberlake}, \citenamefont {Gasbarri}, \citenamefont {Vinante},
  \citenamefont {Setter},\ and\ \citenamefont
  {Ulbricht}}]{timberlake_acceleration_2019}%
  \BibitemOpen
  \bibfield  {author} {\bibinfo {author} {\bibfnamefont {Chris}\ \bibnamefont
  {Timberlake}}, \bibinfo {author} {\bibfnamefont {Giulio}\ \bibnamefont
  {Gasbarri}}, \bibinfo {author} {\bibfnamefont {Andrea}\ \bibnamefont
  {Vinante}}, \bibinfo {author} {\bibfnamefont {Ashley}\ \bibnamefont
  {Setter}}, \ and\ \bibinfo {author} {\bibfnamefont {Hendrik}\ \bibnamefont
  {Ulbricht}},\ }\bibfield  {title} {\enquote {\bibinfo {title} {Acceleration
  sensing with magnetically levitated oscillators above a superconductor},}\
  }\href {\doibase 10.1063/1.5129145} {\bibfield  {journal} {\bibinfo
  {journal} {Appl. Phys. Lett.}\ }\textbf {\bibinfo {volume} {115}},\ \bibinfo
  {pages} {224101} (\bibinfo {year} {2019})}\BibitemShut {NoStop}%
\bibitem [{\citenamefont {Vinante}\ \emph {et~al.}(2020)\citenamefont
  {Vinante}, \citenamefont {Falferi}, \citenamefont {Gasbarri}, \citenamefont
  {Setter}, \citenamefont {Timberlake},\ and\ \citenamefont
  {Ulbricht}}]{vinante_ultralow_2020}%
  \BibitemOpen
  \bibfield  {author} {\bibinfo {author} {\bibfnamefont {A.}~\bibnamefont
  {Vinante}}, \bibinfo {author} {\bibfnamefont {P.}~\bibnamefont {Falferi}},
  \bibinfo {author} {\bibfnamefont {G.}~\bibnamefont {Gasbarri}}, \bibinfo
  {author} {\bibfnamefont {A.}~\bibnamefont {Setter}}, \bibinfo {author}
  {\bibfnamefont {C.}~\bibnamefont {Timberlake}}, \ and\ \bibinfo {author}
  {\bibfnamefont {H.}~\bibnamefont {Ulbricht}},\ }\bibfield  {title} {\enquote
  {\bibinfo {title} {Ultralow {Mechanical} {Damping} with
  {Meissner}-{Levitated} {Ferromagnetic} {Microparticles}},}\ }\href {\doibase
  10.1103/PhysRevApplied.13.064027} {\bibfield  {journal} {\bibinfo  {journal}
  {Phys. Rev. Applied}\ }\textbf {\bibinfo {volume} {13}},\ \bibinfo {pages}
  {064027} (\bibinfo {year} {2020})}\BibitemShut {NoStop}%
\bibitem [{\citenamefont {Gieseler}\ \emph {et~al.}(2020)\citenamefont
  {Gieseler}, \citenamefont {Kabcenell}, \citenamefont {Rosenfeld},
  \citenamefont {Schaefer}, \citenamefont {Safira}, \citenamefont {Schuetz},
  \citenamefont {Gonzalez-Ballestero}, \citenamefont {Rusconi}, \citenamefont
  {Romero-Isart},\ and\ \citenamefont {Lukin}}]{gieseler_single-spin_2020}%
  \BibitemOpen
  \bibfield  {author} {\bibinfo {author} {\bibfnamefont {J.}~\bibnamefont
  {Gieseler}}, \bibinfo {author} {\bibfnamefont {A.}~\bibnamefont {Kabcenell}},
  \bibinfo {author} {\bibfnamefont {E.}~\bibnamefont {Rosenfeld}}, \bibinfo
  {author} {\bibfnamefont {J.~D.}\ \bibnamefont {Schaefer}}, \bibinfo {author}
  {\bibfnamefont {A.}~\bibnamefont {Safira}}, \bibinfo {author} {\bibfnamefont
  {M.~J.~A.}\ \bibnamefont {Schuetz}}, \bibinfo {author} {\bibfnamefont
  {C.}~\bibnamefont {Gonzalez-Ballestero}}, \bibinfo {author} {\bibfnamefont
  {C.~C.}\ \bibnamefont {Rusconi}}, \bibinfo {author} {\bibfnamefont
  {O.}~\bibnamefont {Romero-Isart}}, \ and\ \bibinfo {author} {\bibfnamefont
  {M.~D.}\ \bibnamefont {Lukin}},\ }\bibfield  {title} {\enquote {\bibinfo
  {title} {Single-{Spin} {Magnetomechanics} with {Levitated} {Micromagnets}},}\
  }\href {\doibase 10.1103/PhysRevLett.124.163604} {\bibfield  {journal}
  {\bibinfo  {journal} {Phys. Rev. Lett.}\ }\textbf {\bibinfo {volume} {124}},\
  \bibinfo {pages} {163604} (\bibinfo {year} {2020})}\BibitemShut {NoStop}%
\bibitem [{\citenamefont {Zheng}\ \emph {et~al.}(2020)\citenamefont {Zheng},
  \citenamefont {Leng}, \citenamefont {Kong}, \citenamefont {Li}, \citenamefont
  {Wang}, \citenamefont {Luo}, \citenamefont {Zhao}, \citenamefont {Duan},
  \citenamefont {Huang}, \citenamefont {Du}, \citenamefont {Carlesso},\ and\
  \citenamefont {Bassi}}]{zheng_room_2020}%
  \BibitemOpen
  \bibfield  {author} {\bibinfo {author} {\bibfnamefont {Di}~\bibnamefont
  {Zheng}}, \bibinfo {author} {\bibfnamefont {Yingchun}\ \bibnamefont {Leng}},
  \bibinfo {author} {\bibfnamefont {Xi}~\bibnamefont {Kong}}, \bibinfo {author}
  {\bibfnamefont {Rui}\ \bibnamefont {Li}}, \bibinfo {author} {\bibfnamefont
  {Zizhe}\ \bibnamefont {Wang}}, \bibinfo {author} {\bibfnamefont {Xiaohui}\
  \bibnamefont {Luo}}, \bibinfo {author} {\bibfnamefont {Jie}\ \bibnamefont
  {Zhao}}, \bibinfo {author} {\bibfnamefont {Chang-Kui}\ \bibnamefont {Duan}},
  \bibinfo {author} {\bibfnamefont {Pu}~\bibnamefont {Huang}}, \bibinfo
  {author} {\bibfnamefont {Jiangfeng}\ \bibnamefont {Du}}, \bibinfo {author}
  {\bibfnamefont {Matteo}\ \bibnamefont {Carlesso}}, \ and\ \bibinfo {author}
  {\bibfnamefont {Angelo}\ \bibnamefont {Bassi}},\ }\bibfield  {title}
  {\enquote {\bibinfo {title} {Room temperature test of the continuous
  spontaneous localization model using a levitated micro-oscillator},}\ }\href
  {\doibase 10.1103/PhysRevResearch.2.013057} {\bibfield  {journal} {\bibinfo
  {journal} {Phys. Rev. Research}\ }\textbf {\bibinfo {volume} {2}},\ \bibinfo
  {pages} {013057} (\bibinfo {year} {2020})}\BibitemShut {NoStop}%
\bibitem [{\citenamefont {Simon}\ and\ \citenamefont {Geim}(2000)}]{frog}%
  \BibitemOpen
  \bibfield  {author} {\bibinfo {author} {\bibfnamefont {M.~D.}\ \bibnamefont
  {Simon}}\ and\ \bibinfo {author} {\bibfnamefont {A.~K.}\ \bibnamefont
  {Geim}},\ }\bibfield  {title} {\enquote {\bibinfo {title} {Diamagnetic
  levitation: Flying frogs and floating magnets},}\ }\href {\doibase
  10.1063/1.372654} {\bibfield  {journal} {\bibinfo  {journal} {Journal of
  Applied Physics}\ }\textbf {\bibinfo {volume} {87}},\ \bibinfo {pages}
  {6200--6204} (\bibinfo {year} {2000})}\BibitemShut {NoStop}%
\bibitem [{\citenamefont {Nirrengarten}\ \emph {et~al.}(2006)\citenamefont
  {Nirrengarten}, \citenamefont {Qarry}, \citenamefont {Roux}, \citenamefont
  {Emmert}, \citenamefont {Nogues}, \citenamefont {Brune}, \citenamefont
  {Raimond},\ and\ \citenamefont {Haroche}}]{nirrengarten_realization_2006}%
  \BibitemOpen
  \bibfield  {author} {\bibinfo {author} {\bibfnamefont {T.}~\bibnamefont
  {Nirrengarten}}, \bibinfo {author} {\bibfnamefont {A.}~\bibnamefont {Qarry}},
  \bibinfo {author} {\bibfnamefont {C.}~\bibnamefont {Roux}}, \bibinfo {author}
  {\bibfnamefont {A.}~\bibnamefont {Emmert}}, \bibinfo {author} {\bibfnamefont
  {G.}~\bibnamefont {Nogues}}, \bibinfo {author} {\bibfnamefont
  {M.}~\bibnamefont {Brune}}, \bibinfo {author} {\bibfnamefont {J.-M.}\
  \bibnamefont {Raimond}}, \ and\ \bibinfo {author} {\bibfnamefont
  {S.}~\bibnamefont {Haroche}},\ }\bibfield  {title} {\enquote {\bibinfo
  {title} {Realization of a {Superconducting} {Atom} {Chip}},}\ }\href
  {\doibase 10.1103/PhysRevLett.97.200405} {\bibfield  {journal} {\bibinfo
  {journal} {Phys. Rev. Lett.}\ }\textbf {\bibinfo {volume} {97}},\ \bibinfo
  {pages} {200405} (\bibinfo {year} {2006})}\BibitemShut {NoStop}%
\bibitem [{\citenamefont {Fortágh}\ and\ \citenamefont
  {Zimmermann}(2007)}]{fortagh_magnetic_2007}%
  \BibitemOpen
  \bibfield  {author} {\bibinfo {author} {\bibfnamefont {József}\ \bibnamefont
  {Fortágh}}\ and\ \bibinfo {author} {\bibfnamefont {Claus}\ \bibnamefont
  {Zimmermann}},\ }\bibfield  {title} {\enquote {\bibinfo {title} {Magnetic
  microtraps for ultracold atoms},}\ }\href {\doibase
  10.1103/RevModPhys.79.235} {\bibfield  {journal} {\bibinfo  {journal} {Rev.
  Mod. Phys.}\ }\textbf {\bibinfo {volume} {79}},\ \bibinfo {pages} {235--289}
  (\bibinfo {year} {2007})}\BibitemShut {NoStop}%
\bibitem [{\citenamefont {Dikovsky}\ \emph {et~al.}(2009)\citenamefont
  {Dikovsky}, \citenamefont {Sokolovsky}, \citenamefont {Zhang}, \citenamefont
  {Henkel},\ and\ \citenamefont {Folman}}]{dikovsky_superconducting_2009}%
  \BibitemOpen
  \bibfield  {author} {\bibinfo {author} {\bibfnamefont {V.}~\bibnamefont
  {Dikovsky}}, \bibinfo {author} {\bibfnamefont {V.}~\bibnamefont
  {Sokolovsky}}, \bibinfo {author} {\bibfnamefont {B.}~\bibnamefont {Zhang}},
  \bibinfo {author} {\bibfnamefont {C.}~\bibnamefont {Henkel}}, \ and\ \bibinfo
  {author} {\bibfnamefont {R.}~\bibnamefont {Folman}},\ }\bibfield  {title}
  {\enquote {\bibinfo {title} {Superconducting atom chips: advantages and
  challenges},}\ }\href {\doibase 10.1140/epjd/e2008-00261-5} {\bibfield
  {journal} {\bibinfo  {journal} {The European Physical Journal D}\ }\textbf
  {\bibinfo {volume} {51}},\ \bibinfo {pages} {247--259} (\bibinfo {year}
  {2009})}\BibitemShut {NoStop}%
\bibitem [{\citenamefont {Bernon}\ \emph {et~al.}(2013)\citenamefont {Bernon},
  \citenamefont {Hattermann}, \citenamefont {Bothner}, \citenamefont
  {Knufinke}, \citenamefont {Weiss}, \citenamefont {Jessen}, \citenamefont
  {Cano}, \citenamefont {Kemmler}, \citenamefont {Kleiner}, \citenamefont
  {Koelle},\ and\ \citenamefont {Fortágh}}]{bernon_manipulation_2013}%
  \BibitemOpen
  \bibfield  {author} {\bibinfo {author} {\bibfnamefont {Simon}\ \bibnamefont
  {Bernon}}, \bibinfo {author} {\bibfnamefont {Helge}\ \bibnamefont
  {Hattermann}}, \bibinfo {author} {\bibfnamefont {Daniel}\ \bibnamefont
  {Bothner}}, \bibinfo {author} {\bibfnamefont {Martin}\ \bibnamefont
  {Knufinke}}, \bibinfo {author} {\bibfnamefont {Patrizia}\ \bibnamefont
  {Weiss}}, \bibinfo {author} {\bibfnamefont {Florian}\ \bibnamefont {Jessen}},
  \bibinfo {author} {\bibfnamefont {Daniel}\ \bibnamefont {Cano}}, \bibinfo
  {author} {\bibfnamefont {Matthias}\ \bibnamefont {Kemmler}}, \bibinfo
  {author} {\bibfnamefont {Reinhold}\ \bibnamefont {Kleiner}}, \bibinfo
  {author} {\bibfnamefont {Dieter}\ \bibnamefont {Koelle}}, \ and\ \bibinfo
  {author} {\bibfnamefont {József}\ \bibnamefont {Fortágh}},\ }\bibfield
  {title} {\enquote {\bibinfo {title} {Manipulation and coherence of ultra-cold
  atoms on a superconducting atom chip},}\ }\href {\doibase 10.1038/ncomms3380}
  {\bibfield  {journal} {\bibinfo  {journal} {Nature Communications}\ }\textbf
  {\bibinfo {volume} {4}},\ \bibinfo {pages} {2380} (\bibinfo {year}
  {2013})}\BibitemShut {NoStop}%
\bibitem [{\citenamefont {Hofer}\ and\ \citenamefont
  {Aspelmeyer}(2019)}]{hofer_analytic_2019}%
  \BibitemOpen
  \bibfield  {author} {\bibinfo {author} {\bibfnamefont {J.}~\bibnamefont
  {Hofer}}\ and\ \bibinfo {author} {\bibfnamefont {M.}~\bibnamefont
  {Aspelmeyer}},\ }\bibfield  {title} {\enquote {\bibinfo {title} {Analytic
  solutions to the {Maxwell}–{London} equations and levitation force for a
  superconducting sphere in a quadrupole field},}\ }\href {\doibase
  10.1088/1402-4896/ab0c44} {\bibfield  {journal} {\bibinfo  {journal} {Phys.
  Scr.}\ }\textbf {\bibinfo {volume} {94}},\ \bibinfo {pages} {125508}
  (\bibinfo {year} {2019})}\BibitemShut {NoStop}%
\bibitem [{\citenamefont {Navau}\ and\ \citenamefont
  {Sanchez}(2020)}]{carles_2020}%
  \BibitemOpen
  \bibfield  {author} {\bibinfo {author} {\bibfnamefont {C.}~\bibnamefont
  {Navau}}\ and\ \bibinfo {author} {\bibfnamefont {A.}~\bibnamefont
  {Sanchez}},\ }\href@noop {} {}\bibinfo {howpublished} {private communication}
  (\bibinfo {year} {2020})\BibitemShut {NoStop}%
\bibitem [{\citenamefont {Lin}(2006)}]{lin_theoretical_2006}%
  \BibitemOpen
  \bibfield  {author} {\bibinfo {author} {\bibfnamefont {Qiong-Gui}\
  \bibnamefont {Lin}},\ }\bibfield  {title} {\enquote {\bibinfo {title}
  {Theoretical development of the image method for a general magnetic source in
  the presence of a superconducting sphere or a long superconducting
  cylinder},}\ }\href {\doibase 10.1103/PhysRevB.74.024510} {\bibfield
  {journal} {\bibinfo  {journal} {Phys. Rev. B}\ }\textbf {\bibinfo {volume}
  {74}},\ \bibinfo {pages} {024510} (\bibinfo {year} {2006})}\BibitemShut
  {NoStop}%
\bibitem [{\citenamefont {Kim}\ \emph {et~al.}(2002)\citenamefont {Kim},
  \citenamefont {Hansen}, \citenamefont {Toppari}, \citenamefont {Suppula},\
  and\ \citenamefont {Pekola}}]{pekola}%
  \BibitemOpen
  \bibfield  {author} {\bibinfo {author} {\bibfnamefont {Nam}\ \bibnamefont
  {Kim}}, \bibinfo {author} {\bibfnamefont {Klavs}\ \bibnamefont {Hansen}},
  \bibinfo {author} {\bibfnamefont {Jussi}\ \bibnamefont {Toppari}}, \bibinfo
  {author} {\bibfnamefont {Tarmo}\ \bibnamefont {Suppula}}, \ and\ \bibinfo
  {author} {\bibfnamefont {Jukka}\ \bibnamefont {Pekola}},\ }\bibfield  {title}
  {\enquote {\bibinfo {title} {Fabrication of mesoscopic superconducting nb
  wires using conventional electron-beam lithographic techniques},}\ }\href
  {\doibase 10.1116/1.1445168} {\bibfield  {journal} {\bibinfo  {journal}
  {Journal of Vacuum Science \& Technology B: Microelectronics and Nanometer
  Structures Processing, Measurement, and Phenomena}\ }\textbf {\bibinfo
  {volume} {20}},\ \bibinfo {pages} {386--388} (\bibinfo {year}
  {2002})}\BibitemShut {NoStop}%
\bibitem [{\citenamefont {Cordier}\ \emph
  {et~al.}(1999{\natexlab{a}})\citenamefont {Cordier}, \citenamefont
  {Flament},\ and\ \citenamefont {Dubuc}}]{cordier_3-d_1999}%
  \BibitemOpen
  \bibfield  {author} {\bibinfo {author} {\bibfnamefont {C.}~\bibnamefont
  {Cordier}}, \bibinfo {author} {\bibfnamefont {S.}~\bibnamefont {Flament}}, \
  and\ \bibinfo {author} {\bibfnamefont {C.}~\bibnamefont {Dubuc}},\ }\bibfield
   {title} {\enquote {\bibinfo {title} {A 3-{D} finite element formulation for
  calculating {Meissner} currents in superconductors},}\ }\href {\doibase
  10.1109/77.763249} {\bibfield  {journal} {\bibinfo  {journal} {IEEE
  Transactions on Applied Superconductivity}\ }\textbf {\bibinfo {volume}
  {9}},\ \bibinfo {pages} {2--6} (\bibinfo {year}
  {1999}{\natexlab{a}})}\BibitemShut {NoStop}%
\bibitem [{\citenamefont {Cordier}\ \emph
  {et~al.}(1999{\natexlab{b}})\citenamefont {Cordier}, \citenamefont
  {Flament},\ and\ \citenamefont {Dubuc}}]{cordier_finite-element_1999}%
  \BibitemOpen
  \bibfield  {author} {\bibinfo {author} {\bibfnamefont {C.}~\bibnamefont
  {Cordier}}, \bibinfo {author} {\bibfnamefont {S.}~\bibnamefont {Flament}}, \
  and\ \bibinfo {author} {\bibfnamefont {C.}~\bibnamefont {Dubuc}},\ }\bibfield
   {title} {\enquote {\bibinfo {title} {Finite-element calculation of
  {Meissner} currents in multiply connected superconductors},}\ }\href
  {\doibase 10.1109/77.819341} {\bibfield  {journal} {\bibinfo  {journal} {IEEE
  Transactions on Applied Superconductivity}\ }\textbf {\bibinfo {volume}
  {9}},\ \bibinfo {pages} {4702--4707} (\bibinfo {year}
  {1999}{\natexlab{b}})}\BibitemShut {NoStop}%
\bibitem [{\citenamefont {Grilli}\ \emph {et~al.}(2005)\citenamefont {Grilli},
  \citenamefont {Stavrev}, \citenamefont {Le~Floch}, \citenamefont
  {Costa-Bouzo}, \citenamefont {Vinot}, \citenamefont {Klutsch}, \citenamefont
  {Meunier}, \citenamefont {Tixador},\ and\ \citenamefont
  {Dutoit}}]{grilli_finite-element_2005}%
  \BibitemOpen
  \bibfield  {author} {\bibinfo {author} {\bibfnamefont {F.}~\bibnamefont
  {Grilli}}, \bibinfo {author} {\bibfnamefont {S.}~\bibnamefont {Stavrev}},
  \bibinfo {author} {\bibfnamefont {Y.}~\bibnamefont {Le~Floch}}, \bibinfo
  {author} {\bibfnamefont {M.}~\bibnamefont {Costa-Bouzo}}, \bibinfo {author}
  {\bibfnamefont {E.}~\bibnamefont {Vinot}}, \bibinfo {author} {\bibfnamefont
  {I.}~\bibnamefont {Klutsch}}, \bibinfo {author} {\bibfnamefont
  {G.}~\bibnamefont {Meunier}}, \bibinfo {author} {\bibfnamefont
  {P.}~\bibnamefont {Tixador}}, \ and\ \bibinfo {author} {\bibfnamefont
  {B.}~\bibnamefont {Dutoit}},\ }\bibfield  {title} {\enquote {\bibinfo {title}
  {Finite-element method modeling of superconductors: from 2-{D} to 3-{D}},}\
  }\href {\doibase 10.1109/TASC.2004.839774} {\bibfield  {journal} {\bibinfo
  {journal} {IEEE Transactions on Applied Superconductivity}\ }\textbf
  {\bibinfo {volume} {15}},\ \bibinfo {pages} {17--25} (\bibinfo {year}
  {2005})}\BibitemShut {NoStop}%
\bibitem [{\citenamefont {Campbell}(2011)}]{campbell_introduction_2011}%
  \BibitemOpen
  \bibfield  {author} {\bibinfo {author} {\bibfnamefont {A.~M.}\ \bibnamefont
  {Campbell}},\ }\bibfield  {title} {\enquote {\bibinfo {title} {An
  {Introduction} to {Numerical} {Methods} in {Superconductors}},}\ }\href
  {\doibase 10.1007/s10948-010-0895-5} {\bibfield  {journal} {\bibinfo
  {journal} {Journal of Superconductivity and Novel Magnetism}\ }\textbf
  {\bibinfo {volume} {24}},\ \bibinfo {pages} {27--33} (\bibinfo {year}
  {2011})}\BibitemShut {NoStop}%
\bibitem [{\citenamefont {Simpson}\ \emph {et~al.}(2001)\citenamefont
  {Simpson}, \citenamefont {Lane}, \citenamefont {Immer},\ and\ \citenamefont
  {Youngquist}}]{currentloop}%
  \BibitemOpen
  \bibfield  {author} {\bibinfo {author} {\bibfnamefont {James~C.}\
  \bibnamefont {Simpson}}, \bibinfo {author} {\bibfnamefont {John~E.}\
  \bibnamefont {Lane}}, \bibinfo {author} {\bibfnamefont {Christopher~D.}\
  \bibnamefont {Immer}}, \ and\ \bibinfo {author} {\bibfnamefont {Robert~C.}\
  \bibnamefont {Youngquist}},\ }\bibfield  {title} {\enquote {\bibinfo {title}
  {Simple analytic expressions for the magnetic field of a circular current
  loop},}\ }\href@noop {} {\bibfield  {journal} {\bibinfo  {journal} {NASA
  Technical Reports Server}\ } (\bibinfo {year} {2001})}\BibitemShut {NoStop}%
\bibitem [{\citenamefont {Franssila}(2010)}]{microfab}%
  \BibitemOpen
  \bibfield  {author} {\bibinfo {author} {\bibfnamefont {Sami}\ \bibnamefont
  {Franssila}},\ }\href@noop {} {\emph {\bibinfo {title} {Introduction to
  Microfabrication}}}\ (\bibinfo  {publisher} {John Wiley \& Sons,
  Incorporated},\ \bibinfo {year} {2010})\BibitemShut {NoStop}%
\bibitem [{\citenamefont {Asada}\ and\ \citenamefont
  {Nosé}(1969)}]{asada_superconductivity_1969}%
  \BibitemOpen
  \bibfield  {author} {\bibinfo {author} {\bibfnamefont {Yuji}\ \bibnamefont
  {Asada}}\ and\ \bibinfo {author} {\bibfnamefont {Hiroshi}\ \bibnamefont
  {Nosé}},\ }\bibfield  {title} {\enquote {\bibinfo {title} {Superconductivity
  of {Niobium} {Films}},}\ }\href {\doibase 10.1143/JPSJ.26.347} {\bibfield
  {journal} {\bibinfo  {journal} {J. Phys. Soc. Jpn.}\ }\textbf {\bibinfo
  {volume} {26}},\ \bibinfo {pages} {347--354} (\bibinfo {year}
  {1969})}\BibitemShut {NoStop}%
\bibitem [{\citenamefont {Rusanov}\ \emph {et~al.}(2004)\citenamefont
  {Rusanov}, \citenamefont {Hesselberth},\ and\ \citenamefont
  {Aarts}}]{rusanov_depairing_2004}%
  \BibitemOpen
  \bibfield  {author} {\bibinfo {author} {\bibfnamefont {A.~Yu.}\ \bibnamefont
  {Rusanov}}, \bibinfo {author} {\bibfnamefont {M.~B.~S.}\ \bibnamefont
  {Hesselberth}}, \ and\ \bibinfo {author} {\bibfnamefont {J.}~\bibnamefont
  {Aarts}},\ }\bibfield  {title} {\enquote {\bibinfo {title} {Depairing
  currents in superconducting films of \${\textbackslash}mathrm\{{Nb}\}\$ and
  amorphous \${\textbackslash}mathrm\{{MoGe}\}\$},}\ }\href {\doibase
  10.1103/PhysRevB.70.024510} {\bibfield  {journal} {\bibinfo  {journal} {Phys.
  Rev. B}\ }\textbf {\bibinfo {volume} {70}},\ \bibinfo {pages} {024510}
  (\bibinfo {year} {2004})}\BibitemShut {NoStop}%
\bibitem [{\citenamefont {Kim}\ \emph {et~al.}(2009)\citenamefont {Kim},
  \citenamefont {Kahng}, \citenamefont {Choi},\ and\ \citenamefont
  {Lee}}]{kim_critical_2009}%
  \BibitemOpen
  \bibfield  {author} {\bibinfo {author} {\bibfnamefont {Yun~Won}\ \bibnamefont
  {Kim}}, \bibinfo {author} {\bibfnamefont {Yung~Ho}\ \bibnamefont {Kahng}},
  \bibinfo {author} {\bibfnamefont {Jae-Hyuk}\ \bibnamefont {Choi}}, \ and\
  \bibinfo {author} {\bibfnamefont {Soon-Gul}\ \bibnamefont {Lee}},\ }\bibfield
   {title} {\enquote {\bibinfo {title} {Critical {Properties} of
  {Submicrometer}-{Patterned} {Nb} {Thin} {Film}},}\ }\href {\doibase
  10.1109/TASC.2009.2019099} {\bibfield  {journal} {\bibinfo  {journal} {IEEE
  Transactions on Applied Superconductivity}\ }\textbf {\bibinfo {volume}
  {19}},\ \bibinfo {pages} {2649--2652} (\bibinfo {year} {2009})}\BibitemShut
  {NoStop}%
\bibitem [{\citenamefont {Ricci}\ \emph {et~al.}(2017)\citenamefont {Ricci},
  \citenamefont {Rica}, \citenamefont {Spasenović}, \citenamefont {Gieseler},
  \citenamefont {Rondin}, \citenamefont {Novotny},\ and\ \citenamefont
  {Quidant}}]{ricci_optically_2017}%
  \BibitemOpen
  \bibfield  {author} {\bibinfo {author} {\bibfnamefont {F.}~\bibnamefont
  {Ricci}}, \bibinfo {author} {\bibfnamefont {R.~A.}\ \bibnamefont {Rica}},
  \bibinfo {author} {\bibfnamefont {M.}~\bibnamefont {Spasenović}}, \bibinfo
  {author} {\bibfnamefont {J.}~\bibnamefont {Gieseler}}, \bibinfo {author}
  {\bibfnamefont {L.}~\bibnamefont {Rondin}}, \bibinfo {author} {\bibfnamefont
  {L.}~\bibnamefont {Novotny}}, \ and\ \bibinfo {author} {\bibfnamefont
  {R.}~\bibnamefont {Quidant}},\ }\bibfield  {title} {\enquote {\bibinfo
  {title} {Optically levitated nanoparticle as a model system for stochastic
  bistable dynamics},}\ }\href {\doibase 10.1038/ncomms15141} {\bibfield
  {journal} {\bibinfo  {journal} {Nature Communications}\ }\textbf {\bibinfo
  {volume} {8}},\ \bibinfo {pages} {15141} (\bibinfo {year}
  {2017})}\BibitemShut {NoStop}%
\bibitem [{\citenamefont {Brandt}\ and\ \citenamefont {Clem}(2004)}]{SCrings}%
  \BibitemOpen
  \bibfield  {author} {\bibinfo {author} {\bibfnamefont {Ernst~Helmut}\
  \bibnamefont {Brandt}}\ and\ \bibinfo {author} {\bibfnamefont {John~R.}\
  \bibnamefont {Clem}},\ }\bibfield  {title} {\enquote {\bibinfo {title}
  {Superconducting thin rings with finite penetration depth},}\ }\href
  {\doibase 10.1103/PhysRevB.69.184509} {\bibfield  {journal} {\bibinfo
  {journal} {Phys. Rev. B}\ }\textbf {\bibinfo {volume} {69}},\ \bibinfo
  {pages} {184509} (\bibinfo {year} {2004})}\BibitemShut {NoStop}%
\bibitem [{\citenamefont {Talantsev}\ \emph {et~al.}(2018)\citenamefont
  {Talantsev}, \citenamefont {Pantoja}, \citenamefont {Crump},\ and\
  \citenamefont {Tallon}}]{talantsev_current_2018}%
  \BibitemOpen
  \bibfield  {author} {\bibinfo {author} {\bibfnamefont {E.~F.}\ \bibnamefont
  {Talantsev}}, \bibinfo {author} {\bibfnamefont {A.~E.}\ \bibnamefont
  {Pantoja}}, \bibinfo {author} {\bibfnamefont {W.~P.}\ \bibnamefont {Crump}},
  \ and\ \bibinfo {author} {\bibfnamefont {J.~L.}\ \bibnamefont {Tallon}},\
  }\bibfield  {title} {\enquote {\bibinfo {title} {Current distribution across
  type {II} superconducting films: a new vortex-free critical state},}\ }\href
  {\doibase 10.1038/s41598-018-20279-3} {\bibfield  {journal} {\bibinfo
  {journal} {Scientific Reports}\ }\textbf {\bibinfo {volume} {8}},\ \bibinfo
  {pages} {1--9} (\bibinfo {year} {2018})}\BibitemShut {NoStop}%
\bibitem [{\citenamefont {Bean}(1964)}]{bean_magnetization_1964}%
  \BibitemOpen
  \bibfield  {author} {\bibinfo {author} {\bibfnamefont {Charles~P.}\
  \bibnamefont {Bean}},\ }\bibfield  {title} {\enquote {\bibinfo {title}
  {Magnetization of {High}-{Field} {Superconductors}},}\ }\href {\doibase
  10.1103/RevModPhys.36.31} {\bibfield  {journal} {\bibinfo  {journal} {Rev.
  Mod. Phys.}\ }\textbf {\bibinfo {volume} {36}},\ \bibinfo {pages} {31--39}
  (\bibinfo {year} {1964})}\BibitemShut {NoStop}%
\bibitem [{\citenamefont {Navau}\ \emph {et~al.}(2013)\citenamefont {Navau},
  \citenamefont {Del-Valle},\ and\ \citenamefont
  {Sanchez}}]{navau_macroscopic_2013}%
  \BibitemOpen
  \bibfield  {author} {\bibinfo {author} {\bibfnamefont {C.}~\bibnamefont
  {Navau}}, \bibinfo {author} {\bibfnamefont {N.}~\bibnamefont {Del-Valle}}, \
  and\ \bibinfo {author} {\bibfnamefont {A.}~\bibnamefont {Sanchez}},\
  }\bibfield  {title} {\enquote {\bibinfo {title} {Macroscopic {Modeling} of
  {Magnetization} and {Levitation} of {Hard} {Type}-{II} {Superconductors}:
  {The} {Critical}-{State} {Model}},}\ }\href {\doibase
  10.1109/TASC.2012.2232916} {\bibfield  {journal} {\bibinfo  {journal} {IEEE
  Transactions on Applied Superconductivity}\ }\textbf {\bibinfo {volume}
  {23}},\ \bibinfo {pages} {8201023--8201023} (\bibinfo {year}
  {2013})}\BibitemShut {NoStop}%
\bibitem [{\citenamefont {Via}\ \emph {et~al.}(2014)\citenamefont {Via},
  \citenamefont {Del-Valle}, \citenamefont {Sanchez},\ and\ \citenamefont
  {Navau}}]{via_simultaneous_2014}%
  \BibitemOpen
  \bibfield  {author} {\bibinfo {author} {\bibfnamefont {Guillem}\ \bibnamefont
  {Via}}, \bibinfo {author} {\bibfnamefont {Nuria}\ \bibnamefont {Del-Valle}},
  \bibinfo {author} {\bibfnamefont {Alvaro}\ \bibnamefont {Sanchez}}, \ and\
  \bibinfo {author} {\bibfnamefont {Carles}\ \bibnamefont {Navau}},\ }\bibfield
   {title} {\enquote {\bibinfo {title} {Simultaneous magnetic and transport
  currents in thin film superconductors within the critical-state
  approximation},}\ }\href {\doibase 10.1088/0953-2048/28/1/014003} {\bibfield
  {journal} {\bibinfo  {journal} {Supercond. Sci. Technol.}\ }\textbf {\bibinfo
  {volume} {28}},\ \bibinfo {pages} {014003} (\bibinfo {year}
  {2014})}\BibitemShut {NoStop}%
\bibitem [{\citenamefont {Grilli}\ \emph {et~al.}(2014)\citenamefont {Grilli},
  \citenamefont {Pardo}, \citenamefont {Stenvall}, \citenamefont {Nguyen},
  \citenamefont {Yuan},\ and\ \citenamefont
  {Gömöry}}]{grilli_computation_2014}%
  \BibitemOpen
  \bibfield  {author} {\bibinfo {author} {\bibfnamefont {Francesco}\
  \bibnamefont {Grilli}}, \bibinfo {author} {\bibfnamefont {Enric}\
  \bibnamefont {Pardo}}, \bibinfo {author} {\bibfnamefont {Antti}\ \bibnamefont
  {Stenvall}}, \bibinfo {author} {\bibfnamefont {Doan~N.}\ \bibnamefont
  {Nguyen}}, \bibinfo {author} {\bibfnamefont {Weijia}\ \bibnamefont {Yuan}}, \
  and\ \bibinfo {author} {\bibfnamefont {Fedor}\ \bibnamefont {Gömöry}},\
  }\bibfield  {title} {\enquote {\bibinfo {title} {Computation of {Losses} in
  {HTS} {Under} the {Action} of {Varying} {Magnetic} {Fields} and
  {Currents}},}\ }\href {\doibase 10.1109/TASC.2013.2259827} {\bibfield
  {journal} {\bibinfo  {journal} {IEEE Transactions on Applied
  Superconductivity}\ }\textbf {\bibinfo {volume} {24}},\ \bibinfo {pages}
  {78--110} (\bibinfo {year} {2014})}\BibitemShut {NoStop}%
\bibitem [{\citenamefont {Mykola}\ and\ \citenamefont
  {Fedor}(2019)}]{mykola_v_2019}%
  \BibitemOpen
  \bibfield  {author} {\bibinfo {author} {\bibfnamefont {Solovyov}\
  \bibnamefont {Mykola}}\ and\ \bibinfo {author} {\bibfnamefont {Gömöry}\
  \bibnamefont {Fedor}},\ }\bibfield  {title} {\enquote {\bibinfo {title}
  {A–{V} formulation for numerical modelling of superconductor magnetization
  in true {3D} geometry},}\ }\href {\doibase 10.1088/1361-6668/ab3a85}
  {\bibfield  {journal} {\bibinfo  {journal} {Supercond. Sci. Technol.}\
  }\textbf {\bibinfo {volume} {32}},\ \bibinfo {pages} {115001} (\bibinfo
  {year} {2019})}\BibitemShut {NoStop}%
\bibitem [{\citenamefont {{COMSOL AB, Stockholm, Sweden}}(2019)}]{comsol}%
  \BibitemOpen
  \bibfield  {author} {\bibinfo {author} {\bibnamefont {{COMSOL AB, Stockholm,
  Sweden}}},\ }\href@noop {} {\enquote {\bibinfo {title} {Comsol
  multiphysics},}\ }\bibinfo {howpublished} {\url{www.comsol.com}} (\bibinfo
  {year} {2019})\BibitemShut {NoStop}%
\bibitem [{\citenamefont {Clerk}\ \emph {et~al.}(2010)\citenamefont {Clerk},
  \citenamefont {Devoret}, \citenamefont {Girvin}, \citenamefont {Marquardt},\
  and\ \citenamefont {Schoelkopf}}]{clerk_introduction_2010}%
  \BibitemOpen
  \bibfield  {author} {\bibinfo {author} {\bibfnamefont {A.~A.}\ \bibnamefont
  {Clerk}}, \bibinfo {author} {\bibfnamefont {M.~H.}\ \bibnamefont {Devoret}},
  \bibinfo {author} {\bibfnamefont {S.~M.}\ \bibnamefont {Girvin}}, \bibinfo
  {author} {\bibfnamefont {Florian}\ \bibnamefont {Marquardt}}, \ and\ \bibinfo
  {author} {\bibfnamefont {R.~J.}\ \bibnamefont {Schoelkopf}},\ }\bibfield
  {title} {\enquote {\bibinfo {title} {Introduction to quantum noise,
  measurement, and amplification},}\ }\href {\doibase
  10.1103/RevModPhys.82.1155} {\bibfield  {journal} {\bibinfo  {journal} {Rev.
  Mod. Phys.}\ }\textbf {\bibinfo {volume} {82}},\ \bibinfo {pages}
  {1155--1208} (\bibinfo {year} {2010})}\BibitemShut {NoStop}%
\bibitem [{\citenamefont {Clarke}\ and\ \citenamefont
  {Braginski}(2006)}]{clarke_squid_2006}%
  \BibitemOpen
  \bibfield  {author} {\bibinfo {author} {\bibfnamefont {John}\ \bibnamefont
  {Clarke}}\ and\ \bibinfo {author} {\bibfnamefont {Alex~I.}\ \bibnamefont
  {Braginski}},\ }\href@noop {} {\emph {\bibinfo {title} {The {SQUID}
  {Handbook} {Fundamentals} and {Technology} of {SQUIDs} and {SQUID}
  {Systems}}}},\ Vol.~\bibinfo {volume} {1}\ (\bibinfo  {publisher}
  {Wiley-VCH},\ \bibinfo {address} {Weinheim},\ \bibinfo {year}
  {2006})\BibitemShut {NoStop}%
\bibitem [{\citenamefont {Schurig}(2014)}]{schurig_making_2014}%
  \BibitemOpen
  \bibfield  {author} {\bibinfo {author} {\bibfnamefont {Thomas}\ \bibnamefont
  {Schurig}},\ }\bibfield  {title} {\enquote {\bibinfo {title} {Making {SQUIDs}
  a practical tool for quantum detection and material characterization in the
  micro- and nanoscale},}\ }\href {\doibase 10.1088/1742-6596/568/3/032015}
  {\bibfield  {journal} {\bibinfo  {journal} {Journal of Physics: Conference
  Series}\ }\textbf {\bibinfo {volume} {568}},\ \bibinfo {pages} {032015}
  (\bibinfo {year} {2014})}\BibitemShut {NoStop}%
\bibitem [{\citenamefont {W\"olbing}\ \emph {et~al.}(2013)\citenamefont
  {W\"olbing}, \citenamefont {Nagel}, \citenamefont {Schwarz}, \citenamefont
  {Kieler}, \citenamefont {Weimann}, \citenamefont {Kohlmann}, \citenamefont
  {Zorin}, \citenamefont {Kemmler}, \citenamefont {Kleiner},\ and\
  \citenamefont {Koelle}}]{wolbing_nb_2013}%
  \BibitemOpen
  \bibfield  {author} {\bibinfo {author} {\bibfnamefont {R.}~\bibnamefont
  {W\"olbing}}, \bibinfo {author} {\bibfnamefont {J.}~\bibnamefont {Nagel}},
  \bibinfo {author} {\bibfnamefont {T.}~\bibnamefont {Schwarz}}, \bibinfo
  {author} {\bibfnamefont {O.}~\bibnamefont {Kieler}}, \bibinfo {author}
  {\bibfnamefont {T.}~\bibnamefont {Weimann}}, \bibinfo {author} {\bibfnamefont
  {J.}~\bibnamefont {Kohlmann}}, \bibinfo {author} {\bibfnamefont {A.~B.}\
  \bibnamefont {Zorin}}, \bibinfo {author} {\bibfnamefont {M.}~\bibnamefont
  {Kemmler}}, \bibinfo {author} {\bibfnamefont {R.}~\bibnamefont {Kleiner}}, \
  and\ \bibinfo {author} {\bibfnamefont {D.}~\bibnamefont {Koelle}},\
  }\bibfield  {title} {\enquote {\bibinfo {title} {Nb nano superconducting
  quantum interference devices with high spin sensitivity for operation in
  magnetic fields up to 0.5 {T}},}\ }\href {\doibase 10.1063/1.4804673}
  {\bibfield  {journal} {\bibinfo  {journal} {Appl. Phys. Lett.}\ }\textbf
  {\bibinfo {volume} {102}} (\bibinfo {year} {2013}),\
  10.1063/1.4804673}\BibitemShut {NoStop}%
\bibitem [{\citenamefont {Caputo}\ \emph {et~al.}(2013)\citenamefont {Caputo},
  \citenamefont {Gozzelino}, \citenamefont {Laviano}, \citenamefont {Ghigo},
  \citenamefont {Gerbaldo}, \citenamefont {Noudem}, \citenamefont {Thimont},\
  and\ \citenamefont {Bernstein}}]{caputo_screening_2013}%
  \BibitemOpen
  \bibfield  {author} {\bibinfo {author} {\bibfnamefont {J.-G.}\ \bibnamefont
  {Caputo}}, \bibinfo {author} {\bibfnamefont {L.}~\bibnamefont {Gozzelino}},
  \bibinfo {author} {\bibfnamefont {F.}~\bibnamefont {Laviano}}, \bibinfo
  {author} {\bibfnamefont {G.}~\bibnamefont {Ghigo}}, \bibinfo {author}
  {\bibfnamefont {R.}~\bibnamefont {Gerbaldo}}, \bibinfo {author}
  {\bibfnamefont {J.}~\bibnamefont {Noudem}}, \bibinfo {author} {\bibfnamefont
  {Y.}~\bibnamefont {Thimont}}, \ and\ \bibinfo {author} {\bibfnamefont
  {P.}~\bibnamefont {Bernstein}},\ }\bibfield  {title} {\enquote {\bibinfo
  {title} {Screening magnetic fields by superconductors: {A} simple model},}\
  }\href {\doibase 10.1063/1.4848015} {\bibfield  {journal} {\bibinfo
  {journal} {Journal of Applied Physics}\ }\textbf {\bibinfo {volume} {114}},\
  \bibinfo {pages} {233913} (\bibinfo {year} {2013})}\BibitemShut {NoStop}%
\bibitem [{\citenamefont {Niepce}\ \emph {et~al.}(2020)\citenamefont {Niepce},
  \citenamefont {Burnett}, \citenamefont {Latorre},\ and\ \citenamefont
  {Bylander}}]{niepce_geometric_2020}%
  \BibitemOpen
  \bibfield  {author} {\bibinfo {author} {\bibfnamefont {David}\ \bibnamefont
  {Niepce}}, \bibinfo {author} {\bibfnamefont {Jonathan~J.}\ \bibnamefont
  {Burnett}}, \bibinfo {author} {\bibfnamefont {Martí~Gutierrez}\ \bibnamefont
  {Latorre}}, \ and\ \bibinfo {author} {\bibfnamefont {Jonas}\ \bibnamefont
  {Bylander}},\ }\bibfield  {title} {\enquote {\bibinfo {title} {Geometric
  scaling of two-level-system loss in superconducting resonators},}\ }\href
  {\doibase 10.1088/1361-6668/ab6179} {\bibfield  {journal} {\bibinfo
  {journal} {Supercond. Sci. Technol.}\ }\textbf {\bibinfo {volume} {33}},\
  \bibinfo {pages} {025013} (\bibinfo {year} {2020})}\BibitemShut {NoStop}%
\bibitem [{\citenamefont {Rosenberg}\ \emph {et~al.}(2017)\citenamefont
  {Rosenberg}, \citenamefont {Kim}, \citenamefont {Das}, \citenamefont {Yost},
  \citenamefont {Gustavsson}, \citenamefont {Hover}, \citenamefont {Krantz},
  \citenamefont {Melville}, \citenamefont {Racz}, \citenamefont {Samach},
  \citenamefont {Weber}, \citenamefont {Yan}, \citenamefont {Yoder},
  \citenamefont {Kerman},\ and\ \citenamefont {Oliver}}]{rosenberg_3d_2017}%
  \BibitemOpen
  \bibfield  {author} {\bibinfo {author} {\bibfnamefont {D.}~\bibnamefont
  {Rosenberg}}, \bibinfo {author} {\bibfnamefont {D.}~\bibnamefont {Kim}},
  \bibinfo {author} {\bibfnamefont {R.}~\bibnamefont {Das}}, \bibinfo {author}
  {\bibfnamefont {D.}~\bibnamefont {Yost}}, \bibinfo {author} {\bibfnamefont
  {S.}~\bibnamefont {Gustavsson}}, \bibinfo {author} {\bibfnamefont
  {D.}~\bibnamefont {Hover}}, \bibinfo {author} {\bibfnamefont
  {P.}~\bibnamefont {Krantz}}, \bibinfo {author} {\bibfnamefont
  {A.}~\bibnamefont {Melville}}, \bibinfo {author} {\bibfnamefont
  {L.}~\bibnamefont {Racz}}, \bibinfo {author} {\bibfnamefont {G.~O.}\
  \bibnamefont {Samach}}, \bibinfo {author} {\bibfnamefont {S.~J.}\
  \bibnamefont {Weber}}, \bibinfo {author} {\bibfnamefont {F.}~\bibnamefont
  {Yan}}, \bibinfo {author} {\bibfnamefont {J.~L.}\ \bibnamefont {Yoder}},
  \bibinfo {author} {\bibfnamefont {A.~J.}\ \bibnamefont {Kerman}}, \ and\
  \bibinfo {author} {\bibfnamefont {W.~D.}\ \bibnamefont {Oliver}},\ }\bibfield
   {title} {\enquote {\bibinfo {title} {{3D} integrated superconducting
  qubits},}\ }\href {\doibase 10.1038/s41534-017-0044-0} {\bibfield  {journal}
  {\bibinfo  {journal} {npj Quantum Information}\ }\textbf {\bibinfo {volume}
  {3}},\ \bibinfo {pages} {1--5} (\bibinfo {year} {2017})}\BibitemShut
  {NoStop}%
\bibitem [{\citenamefont {Grilli}\ \emph {et~al.}(2013)\citenamefont {Grilli},
  \citenamefont {Brambilla}, \citenamefont {Sirois}, \citenamefont {Stenvall},\
  and\ \citenamefont {Memiaghe}}]{grilli_development_2013}%
  \BibitemOpen
  \bibfield  {author} {\bibinfo {author} {\bibfnamefont {Francesco}\
  \bibnamefont {Grilli}}, \bibinfo {author} {\bibfnamefont {Roberto}\
  \bibnamefont {Brambilla}}, \bibinfo {author} {\bibfnamefont {Frédéric}\
  \bibnamefont {Sirois}}, \bibinfo {author} {\bibfnamefont {Antti}\
  \bibnamefont {Stenvall}}, \ and\ \bibinfo {author} {\bibfnamefont {Steeve}\
  \bibnamefont {Memiaghe}},\ }\bibfield  {title} {\enquote {\bibinfo {title}
  {Development of a three-dimensional finite-element model for high-temperature
  superconductors based on the {H}-formulation},}\ }\href {\doibase
  10.1016/j.cryogenics.2012.03.007} {\bibfield  {journal} {\bibinfo  {journal}
  {Cryogenics}\ }\textbf {\bibinfo {volume} {53}},\ \bibinfo {pages} {142--147}
  (\bibinfo {year} {2013})}\BibitemShut {NoStop}%
\bibitem [{\citenamefont {Morandi}\ \emph {et~al.}(2017)\citenamefont
  {Morandi}, \citenamefont {Ainslie}, \citenamefont {Grilli},\ and\
  \citenamefont {Stenvall}}]{morandi_5th_2017}%
  \BibitemOpen
  \bibfield  {author} {\bibinfo {author} {\bibfnamefont {Antonio}\ \bibnamefont
  {Morandi}}, \bibinfo {author} {\bibfnamefont {Mark~D.}\ \bibnamefont
  {Ainslie}}, \bibinfo {author} {\bibfnamefont {Francesco}\ \bibnamefont
  {Grilli}}, \ and\ \bibinfo {author} {\bibfnamefont {Antti}\ \bibnamefont
  {Stenvall}},\ }\bibfield  {title} {\enquote {\bibinfo {title} {The 5th
  international workshop on numerical modelling of high temperature
  superconductors},}\ }\href {\doibase 10.1088/1361-6668/aa7676} {\bibfield
  {journal} {\bibinfo  {journal} {Supercond. Sci. Technol.}\ }\textbf {\bibinfo
  {volume} {30}},\ \bibinfo {pages} {080201} (\bibinfo {year}
  {2017})}\BibitemShut {NoStop}%
\bibitem [{\citenamefont {Grilli}\ \emph {et~al.}(2018)\citenamefont {Grilli},
  \citenamefont {Morandi}, \citenamefont {Silvestri},\ and\ \citenamefont
  {Brambilla}}]{grilli_dynamic_2018}%
  \BibitemOpen
  \bibfield  {author} {\bibinfo {author} {\bibfnamefont {Francesco}\
  \bibnamefont {Grilli}}, \bibinfo {author} {\bibfnamefont {Antonio}\
  \bibnamefont {Morandi}}, \bibinfo {author} {\bibfnamefont {Federica~De}\
  \bibnamefont {Silvestri}}, \ and\ \bibinfo {author} {\bibfnamefont {Roberto}\
  \bibnamefont {Brambilla}},\ }\bibfield  {title} {\enquote {\bibinfo {title}
  {Dynamic modeling of levitation of a superconducting bulk by coupled
  {H}-magnetic field and arbitrary {Lagrangian}–{Eulerian} formulations},}\
  }\href {\doibase 10.1088/1361-6668/aae426} {\bibfield  {journal} {\bibinfo
  {journal} {Supercond. Sci. Technol.}\ }\textbf {\bibinfo {volume} {31}},\
  \bibinfo {pages} {125003} (\bibinfo {year} {2018})}\BibitemShut {NoStop}%
\bibitem [{\citenamefont {Kovetz}(1990)}]{kovetz1990principles}%
  \BibitemOpen
  \bibfield  {author} {\bibinfo {author} {\bibfnamefont {Attay}\ \bibnamefont
  {Kovetz}},\ }\href@noop {} {\emph {\bibinfo {title} {The principles of
  electromagnetic theory}}}\ (\bibinfo  {publisher} {CUP Archive},\ \bibinfo
  {year} {1990})\BibitemShut {NoStop}%
\bibitem [{\citenamefont {Tinkham}(2004)}]{Tinkham}%
  \BibitemOpen
  \bibfield  {author} {\bibinfo {author} {\bibfnamefont {Michael}\ \bibnamefont
  {Tinkham}},\ }\href@noop {} {\emph {\bibinfo {title} {Introduction to
  superconductivity}}}\ (\bibinfo  {publisher} {Dover publ.},\ \bibinfo {year}
  {2004})\BibitemShut {NoStop}%
\bibitem [{\citenamefont {Annett}(2004)}]{anett_superconductivity}%
  \BibitemOpen
  \bibfield  {author} {\bibinfo {author} {\bibfnamefont {James~F.}\
  \bibnamefont {Annett}},\ }\href@noop {} {\emph {\bibinfo {title}
  {Superconductivity, superfluids, and condensates}}},\ Vol.~\bibinfo {volume}
  {5}\ (\bibinfo  {publisher} {Oxford University Press},\ \bibinfo {address}
  {New York},\ \bibinfo {year} {2004})\BibitemShut {NoStop}%
\bibitem [{\citenamefont {Beleggia}\ \emph {et~al.}(2009)\citenamefont
  {Beleggia}, \citenamefont {Vokoun},\ and\ \citenamefont {Graef}}]{demag}%
  \BibitemOpen
  \bibfield  {author} {\bibinfo {author} {\bibfnamefont {M.}~\bibnamefont
  {Beleggia}}, \bibinfo {author} {\bibfnamefont {D.}~\bibnamefont {Vokoun}}, \
  and\ \bibinfo {author} {\bibfnamefont {M.~De}\ \bibnamefont {Graef}},\
  }\bibfield  {title} {\enquote {\bibinfo {title} {Demagnetization factors for
  cylindrical shells and related shapes},}\ }\href {\doibase
  https://doi.org/10.1016/j.jmmm.2008.11.046} {\bibfield  {journal} {\bibinfo
  {journal} {Journal of Magnetism and Magnetic Materials}\ }\textbf {\bibinfo
  {volume} {321}},\ \bibinfo {pages} {1306 -- 1315} (\bibinfo {year}
  {2009})}\BibitemShut {NoStop}%
\end{thebibliography}%

\end{document}